\documentclass[journal, 10pt]{IEEEtran}

\usepackage[cmex10]{amsmath}
\usepackage{amssymb}
\usepackage{amsfonts}
\usepackage{graphics} 
\usepackage{epsfig}
\usepackage{epstopdf}
\usepackage{subfigure}  
\usepackage{color}
\usepackage{cite}
\usepackage{graphicx}
\usepackage{diagbox}
\usepackage{booktabs}
\usepackage{stfloats}

\allowdisplaybreaks[4]

\newtheorem{remark}{Remark}

\ifCLASSINFOpdf
\else
\fi

\begin{document}

\title{Joint Long-Term Admission Control and Beamforming in Green Downlink Networks: Offline and Online Approaches}

\author{Jingran Lin, Mengyuan Ma, Qiang Li, and Jian Yang
\thanks{
Part of this paper has been presented at EUSIPCO~2018 \cite{Lin2018C}.}
\thanks{J. Lin, M. Ma, and Q. Li are with the School of Information and Communication Engineering, University of Electronic Science and Technology of China, Chengdu 611731, Sichuan, P.R. China (e-mail: jingranlin@uestc.edu.cn; mamengyuan@std.uestc.edu.cn; lq@uestc.edu.cn).}
\thanks{J. Yang is with the Northern Institute of Electronic Equipment of China, Beijing 100191, P.R. China (e-mail: jyang\_niee@outlook.com).}
}

\markboth{DRAFT}{}
%

\maketitle

\begin{abstract}
  Admission control is an effective solution to managing a wireless network with a large user set. It dynamically rejects users in bad conditions, and thereby guarantees high quality-of-service~(QoS) for the others. Unfortunately, a frequently-varying admissible user set requests remarkable power to continually re-establish the transmission link. Hence, we explore the stability of admissible user set, and formulate a joint long-term admission control and beamforming problem for a network with one multi-antenna base station~(BS) and multiple single-antenna users. We consider the downlink transmission in a time period containing multiple time slices. By jointly optimizing the admissible users and the BS transmit beamformers in different time slices, we aim to minimize the total power cost, including not only the power for QoS guarantee, but also the power for user status switching. Due to the NP-hardness of the problem, we develop two (offline and online) algorithms to find some efficient suboptimal solutions. The offline algorithm requires all channel state information~(CSI), and solves the admissible users and beamformers in different time slices in one shot, based on successive upper-bound minimization (SUM). To support real-time data transmission, we further design an alternating direction method of multipliers~(ADMM)-based online algorithm, which outputs the admissible users and beamformers in different time slices successively, utilizing the previous admissible user set, the actual value of current CSI, and the distribution of future CSI. Simulations validate the performance of the two algorithms, and show that the online algorithm is an efficient practical alternative to the offline algorithm.
\end{abstract}


\IEEEpeerreviewmaketitle

\section{Introduction} \label{sec:introduction}
\IEEEPARstart{I}{n} general, the quality-of-service~(QoS) of a wireless network degrades rapidly with the size of user set, due to the increased competition among users. Admission control \cite{Ahmed2005J} is an effective approach to this problem, which dynamically rejects the users in bad conditions to guarantee high QoS for the rest. Currently, admission control has been widely employed in wireless networks to balance the power cost and the size of user set \cite{Andersin1996J, Mitliagkas2011J, Liu2013J, Liu2015J, Matskani2008J, Lin2017J, Wai2012C, Evangelinakis2010C}, or balance the the spectral/energy efficiencies and the size of user set \cite{Lai2016J, Kuang2012C, Azam2016J, Liu2012C, Monemi2015J}. In fact, the admission control problems are not easy (usually shown to be NP-hard \cite{Andersin1996J, Mitliagkas2011J, Matskani2008J}). As a compromise, many suboptimal approaches, e.g., deflation heuristic \cite{Zander1992J1, Andersin1996J, Lin2017C1}, convex approximation \cite{Mitliagkas2011J, Liu2013J, Lin2017J, Lin2019J, Matskani2008J, Wai2012C,Lai2016J, Kuang2012C}, non-convex ($\ell_q$) approximation \cite{Liu2015J,Lai2016J}, etc., have been designed to find efficient suboptimal solutions.

\begin{figure}[tbp]
  \centering
  \includegraphics[width = 7cm]{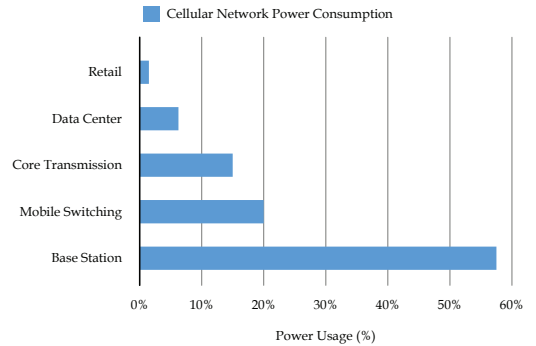}
  \caption{Power consumption of a typical wireless cellular network \cite{Han2011J,Hasan2011J}.}
  \label{fig:power_usage}
\end{figure}

On the other hand, most current studies, though different in scenario setting and solution approach, optimize the admissible users based on \emph{instantaneous} channel state information (CSI). Notice that the fading characteristics in radio propagation lead to time-varying wireless channels. In consequence, the admissible status of each user may switch between ``admissible'' and ``inadmissible'' frequently, thus yielding many practical issues in communication management. For instance, when the user is admitted by the network, some necessary yet complicated operations, e.g., channel estimation, synchronization, handover, etc., need to be performed in order to establish the transmission link between the base station (BS) and the user. Consequently, a frequently-varying admissible user set causes non-negligible signaling and power cost \cite{Chen2011J, Han2011J, Hasan2011J}. As shown in Fig. \ref{fig:power_usage}, about 20\% of the total power cost in a typical wireless network is consumed for transmission link switching, which even exceeds that for data transmission. Moreover, if the admissible status of a user changes continually, it may suffer from serious service break due to the frequent interruptions in data transmission. In view of this, we are motivated to balance the flexibility and stability of user set in admission control, which thereby yields the problem of \emph{long-term} admission control.

\subsection{Related Works}
Essentially, long-term admission control falls into the category of problems exploring the stability of transmission links. In early studies, the hysteresis margin is widely used to avoid frequent user switching \cite{Simsek2015C}, where the user does not switch to another BS until the difference between the required QoS and the achievable QoS exceeds the hysteresis margin. Obviously, this approach improves the stability of transmission links at the cost of QoS. In \cite{Pan2012C}, by employing the Markov decision processes (MDP), a switching algorithm considering transmission latency and handover signaling cost is developed. The authors of \cite{Lin2017C2,Sun2013C,Yu2016J} address this problem from the perspective of BS activation, where they choose the active BSs in different time slices properly to control the switching frequency of each user among distinct BSs. Unfortunately, these approaches cannot be directly applied to our problem, where we pursue relatively stable transmission links by selecting users (instead of BSs). The authors of \cite{Chen2017J} utilize the channel distribution information (CDI) and optimize the admissible users by solving an outage-constrained problem. In this design, the admissible user set is fixed as long as the CDI does not change. However, there is some risk that the QoS requests of the admissible users cannot be satisfied due to the randomly-varying channels.

\subsection{Contributions of Our Work}
In this paper, we consider the green communication problem in a downlink network with one multi-antenna BS and multiple single-antenna users. We aim to balance the size of user set and the power cost. Different from most current studies minimizing the power for data transmission only, we further take the power for transmission link switching into account. To this target, we propose the problem of joint long-term admission control and beamforming, and design efficient algorithms for it. Our main contributions are summarized as follows.

\begin{enumerate}
  \item We propose a formulation of joint long-term admission control and beamforming for green communication. We consider the downlink transmission in a time period consisting of multiple time slices. Under the QoS and power budget constraints, we jointly optimize the admissible users and the BS transmit beamformers in different time slices, such that the size of user set, the transmit power, and the switching power can be well balanced to reduce the total system cost.
  \item To handle this challenging problem, we first develop an offline algorithm for it, where we assume the availability of all CSIs in the time period, and solve the variables in different time slices in one shot. Concretely, the problem is solved based on successive upper-bound minimization (SUM) \cite{Meisam2013J}, which sequentially approximates the non-convex and non-smooth objective by a global tight upper bound, and finally achieves a stationary solution of the problem.
  \item To support real-time data transmission, we further design an online approach to the problem, which optimizes the admissible users and the BS transmit beamformers in different time slices successively, utilizing the previous admissible user set, the actual value of current CSI, and the distribution of future CSI. To avoid solving a difficult stochastic problem, we employ the sample average approximation (SAA) method \cite{Chen2017J, Shapiro2006J}, and then obtain a deterministic problem. Next, an alternating direction method of multipliers (ADMM)-based algorithm \cite{Boyd2011J,Shen2012J} is designed to efficiently solve the admission control and beamforming problem in each time slice. Finally, a low-complexity online algorithm is proposed.
\end{enumerate}

\subsection{Organization and Notations}
The rest of this paper is organized as follows. The system model and problem statement are given in Section \ref{sec:model}. In Sections \ref{sec:offline} and \ref{sec:online}, we develop the offline and online algorithms for the problem, respectively. Simulation results are shown in Section \ref{sec:simulation}. Section \ref{sec:conclusion} concludes this paper.

\emph{Notations:} We use $(\cdot)^T$ and $(\cdot)^\dag$ to denote the transpose and Hermitian of a matrix (or vector), respectively; $(\cdot)^{-1}$ and ${\rm Tr}(\cdot)$ denote the inverse and the trace of a matrix. Denote $\|\cdot\|_p$ and $\|\cdot\|_F$ as the $\ell_p$-norm $(p = 0, 1, 2)$ and the Frobenius norm. $\mathbb{R}^{M \times N}$ (or $\mathbb{C}^{M \times N}$) is the set of all $M \times N$ real (or complex) matrices. ${\Re}\{\cdot\}$ and ${\Im}\{\cdot\}$ denote the real part and the imaginary part of a complex number, respectively. $[\cdot]_x^+ \triangleq \max\{\cdot, x\}$ with $x$ being a real number. $\mathbb{E}\{\cdot\}$ denotes expectation.

\section{System Model and Problem Statement} \label{sec:model}
We consider a downlink network where one $N$-antenna BS provides service to $M$ single-antenna users. The block fading channel model \cite{Biglieri1998J} is employed here --- the channels remain static in each fading block, while changing randomly from one block to another according to certain channel distribution. We define each fading block as a time slice, and consider the data transmission in a time period comprised of $T$ time slices. Let $\mathbf{w}_m(t) \in \mathbb{C}^{N \times 1}$ and $\mathbf{h}_m(t) \in \mathbb{C}^{N \times 1}$ denote the BS transmit beamformer and the channel vector of user $m$ in time slice $t$, for $m = 1, 2, \ldots, M$, and $t = 1,2, \ldots, T$. We further define $\mathbf{W}(t) \triangleq [\mathbf{w}_1(t), \mathbf{w}_2(t), \ldots, \mathbf{w}_M(t)]\in \mathbb{C}^{N \times M}$ and $\mathbf{H}(t) \triangleq  [\mathbf{h}_1(t), \mathbf{h}_2(t), \ldots, \mathbf{h}_M(t)] \in \mathbb{C}^{N \times M}$ as the beamformer matrix and channel matrix of all users in time slice $t$, $t = 1, 2, \ldots, T$.

\subsection{Conventional Joint Admission Control and Beamforming}
Firstly, let us take a review of the conventional QoS-based joint admission control and beamforming problem. We use the signal-to-interference-plus-noise ratio~(SINR) to evaluate the QoS of each user. For our system model, the SINR of user $m$ in time slice $t$ is given by
\begin{equation}
    {\rm SINR}_m(t) = \tfrac{|\mathbf{h}_m^\dag(t)\mathbf{w}_m(t)|^2}{\sigma^2 + \sum_{n \neq m}|\mathbf{h}_m^\dag(t)\mathbf{w}_n(t)|^2},~\forall\;m,t, \label{eq:SINR}
\end{equation}
where $\sigma^2$ is the noise power. Define $\gamma$ as the desired QoS level, and then we admit users according to whether the QoS constraints, i.e., ${\rm SINR}_m(t) \geq \gamma,~ \forall \, m, t$, can be satisfied. Applying the second-order cone programming (SOCP) reformulation \cite{Matskani2008J} and introducing a series of variables $v_m(t) \geq 0,\forall\, m,t$, we get a reformulation of ${\rm SINR}_m(t) \geq \gamma$, i.e.,
\begin{equation}
    \left\{\begin{aligned}
        &\tfrac{\mathbf{h}_m^\dag(t)\mathbf{w}_m(t) + v_m(t)}{\sqrt{\sigma^2 + \sum_{n \neq m}|\mathbf{h}_m^\dag(t)\mathbf{w}_n(t)|^2}} \geq \sqrt{\gamma}, \\
        &{\Im}\{\mathbf{h}_m^\dag(t)\mathbf{w}_m(t)\} = 0.
    \end{aligned}\right.\label{cstr:QoS_SOCP}
\end{equation}
Actually, $v_m(t)$ works as the admissible indicator of user $m$ in time slice $t$. Specifically, $v_m(t) = 0$ indicates that user $m$ can be served at its desired QoS level $\gamma$, and hence is admissible, while $v_m(t) > 0$ indicates that user $m$ should be rejected.

Hence, the conventional QoS-based joint admission control and beamforming problem in time slice $t$ is formulated as
\begin{subequations} \label{prob:conventional}
\begin{align}
    \min_{\{\mathbf{v}(t), \mathbf{W}(t)\}} ~& \|\mathbf{W}(t)\|_F^2 + \lambda_1 \|\mathbf{v}(t)\|_0 \nonumber \\
    {\rm s.t.}\qquad & \eqref{cstr:QoS_SOCP}~{\rm is~satisfied},~\forall \, m, \nonumber \\
    & \|\mathbf{W}(t)\|_F^2 \leq P, \label{cstr:conventional_power} \\
    & v_m(t) \geq 0,~ \forall\,m, \label{cstr:conventional_indicator}
\end{align}
\end{subequations}
where $\mathbf{v}(t) = [v_1(t), v_2(t), \ldots, v_M(t)] \in \mathbb{R}^{1 \times M}$, and $\|\mathbf{v}(t)\|_0$ denotes the number of inadmissible users in time slice $t$; the weighting factor $\lambda_1$ can be viewed as the cost (profit loss) of rejecting one user
; $P$ is the BS transmit power budget.

In problem~\eqref{prob:conventional}, we optimize the admissible users and the BS transmit beamformers in different time slices independently, based on instantaneous CSI $\mathbf{H}(t)$. Since $\mathbf{H}(t)$ changes with $t$ randomly according to the channel distribution, the user status may keep switching between admissible and inadmissible, thus requesting considerable power cost to frequently re-establish the transmission link. To alleviate this, we address the stability of admissible user set in the following discussion.

\subsection{Joint Long-Term Admission Control and Beamforming}
To address the stability of each user's admissible status, we define the following function $\mathcal{I}(\cdot)$, i.e.,
\begin{equation} \label{eq:indicator_func}
    \mathcal{I}(x) = \left\{\begin{aligned}
                        &0, ~~ x = 0, \quad{\rm{(admissible)}}\\
                        &1, ~~ x > 0, \quad{\rm{(inadmissible)}}
                     \end{aligned}\right.
\end{equation}
to map $v_m(t)$ to a binary admissible status. It is easy to show that the number of inadmissible users in time slice $t$ can be described by $\sum_{m = 1}^M \mathcal{I}[v_m(t)]$, since we have
\begin{equation}
    \|\mathbf{v}(t)\|_0 = \sum\limits_{m = 1}^M \mathcal{I}[v_m(t)],~\forall \, t.
\end{equation}
Further, $\mathcal{I}[v_m(t+1)] - \mathcal{I}[v_m(t)] = 0$ means that the admissible status of user $m$ does not change from time slice $t$ to time slice $(t+1)$; otherwise, $\mathcal{I}[v_m(t+1)] - \mathcal{I}[v_m(t)] \neq 0$ means that the admissible status changes\footnote{We cannot directly use $[v_m(t+1) - v_m(t)]$ to indicate the variation of user $m$'s status. For instance, in the case of $v_m(t) > 0$, $v_m(t+1) > 0$, and $v_m(t) \neq v_m(t+1)$, the user status does not change (keeps inadmissible in the adjacent two time slices), while we have $v_m(t+1) - v_m(t) \neq 0$.}. To limit the frequent switching of each user's admissible status, we should optimize the values of $\mathcal{I}[v_m(t+1)] - \mathcal{I}[v_m(t)]$, $\forall \, m, t$. This drives us to consider the admission control problems in different time slices jointly, thus leading to the following joint long-term admission control and beamforming problem
\begin{subequations} \label{prob:original_prob}
\begin{align}
    \min_{\{\mathbf{v}, \mathbf{W}\}} ~&\left\{\begin{aligned}
        &\sum_{t=1}^T\|\mathbf{W}(t)\|_F^2 + \lambda_1\sum_{t=1}^T\sum_{m=1}^M\mathcal{I}[v_m(t)]\\
        &+\lambda_2\sum_{t=1}^{T-1}\sum_{m=1}^M \left|\mathcal{I}[v_m(t+1)] - \mathcal{I}[v_m(t)]\right|
    \end{aligned}\right\} \label{obj:original_prob} \\
    {\rm s.t.}\quad & \eqref{cstr:QoS_SOCP}~{\rm is~satisfied},~\forall\,m,t, \nonumber \\
        &\|\mathbf{W}(t)\|_F^2 \leq P,~\forall\,t, \label{cstr:power} \\
        &v_m(t) \geq 0,~ \forall\,m,t, \label{cstr:indicator}
\end{align}
\end{subequations}
where $\{\mathbf{v}, \mathbf{W}\}$ denotes $\{\mathbf{v}(t), \mathbf{W}(t)\}_{t = 1}^T$; $\lambda_2$ is the power cost to support one transmission link switching.

In problem \eqref{prob:original_prob}, by jointly optimizing the admissible users and the BS transmit beamformers in the length-$T$ time period, we aim to balance the size of admissible user set, the transmit power, and the switching power. It should be mentioned that instead of forcing an absolutely static user set \cite{Chen2017J}, we seek a relatively stable user set by limiting the switching frequency of each user's admissible status --- here the user set is allowed to change, but its variation should be carefully controlled.
\newcounter{mytempeqncnt}
\begin{figure*}[b]
  \hrulefill
  \setcounter{mytempeqncnt}{\value{equation}}
  \setcounter{equation}{9}
  \begin{equation} \label{eq:upper_bound_objective}
    u(\mathbf{v};\bar{\mathbf{v}}) \triangleq \lambda_1\sum\limits_{t=1}^T\sum\limits_{m=1}^M \! \left[ 1 - \tfrac{2}{1+\kappa \bar{v}_m(t)} + \tfrac{1+\kappa v_m(t)}{[1+\kappa \bar{v}_m(t)]^2}\right]
        + \lambda_2\sum\limits_{t=1}^{T-1}\sum\limits_{m=1}^M \max\!\left\{\begin{aligned}
        &\! \tfrac{1}{1+\kappa v_m(t+1)} - \tfrac{2}{1+\kappa\bar{v}_m(t)} + \tfrac{1+\kappa v_m(t)}{[1+\kappa\bar{v}_m(t)]^2},\\
        &\! \tfrac{1}{1+\kappa v_m(t)} - \tfrac{2}{1+\kappa\bar{v}_m(t+1)} + \tfrac{1+\kappa v_m(t+1)}{[1+\kappa\bar{v}_m(t+1)]^2}
    \end{aligned}\right\},
  \end{equation}
  \setcounter{equation}{12}
  \begin{align}
    p_c(t) &\triangleq \|\mathbf{W}(t)\|_F^2 + \lambda_1\sum\limits_{m = 1}^M \! \left[1 - \tfrac{1}{1+\kappa v_m(t)}\right] + \lambda_2 \sum\limits_{m=1}^M \! \left|\tfrac{1}{1+\kappa v_m(t-1)} - \tfrac{1}{1+\kappa v_m(t)}\right| \nonumber \\
    &\qquad\qquad~~~ + \mathbb{E} \left\{\|\mathbf{W}(t+1)\|_F^2 + \lambda_1\sum\limits_{m = 1}^M \! \left[1 - \tfrac{1}{1+\kappa v_m(t+1)}\right] + \lambda_2 \sum\limits_{m=1}^M \! \left|\tfrac{1}{1+\kappa v_m(t)} - \tfrac{1}{1+\kappa v_m(t+1)}\right|\right\}, \qquad\qquad~\,\,\label{eq:power_cost_t}
  \end{align}
  \begin{align}
    \hat{p}_c(t) &\triangleq \|\mathbf{W}(t)\|_F^2 + \lambda_1\sum\limits_{m = 1}^M \left[1 - \tfrac{1}{1+\kappa v_m(t)}\right] + \lambda_2 \sum\limits_{m=1}^M \left|\tfrac{1}{1+\kappa v_m(t-1)} - \tfrac{1}{1+\kappa v_m(t)}\right| \nonumber \\
    &\qquad\qquad~~~ + \frac{1}{J}\sum\limits_{j=1}^J \left\{\|\hat{\mathbf{W}}(t+1,j)\|_F^2 + \lambda_1\sum\limits_{m = 1}^M \left[1 - \tfrac{1}{1+\kappa \hat{v}_m(t+1,j)}\right] + \lambda_2 \sum\limits_{m=1}^M \left|\tfrac{1}{1+\kappa v_m(t)} - \tfrac{1}{1+\kappa \hat{v}_m(t+1,j)}\right|\right\},~ \label{eq:power_cost_t_saa}
  \end{align}
  \setcounter{equation}{\value{mytempeqncnt}}
\end{figure*}
\setcounter{equation}{6}

To tackle the challenging (essentially NP-hard) problem \eqref{prob:original_prob}, we first design an SUM-based offline algorithm to solve the problem in one shot, by assuming the availability of all CSIs within the time period. Next, in order to support real-time data transmission, we further design an online algorithm which solves the problem time slice by time slice.

\section{Offline Algorithm for Problem~\eqref{prob:original_prob}} \label{sec:offline}
Obviously, the non-continuous function $\mathcal{I}(\cdot)$ makes problem \eqref{prob:original_prob} difficult. To bypass this, we apply the approximation of
\begin{equation}
    \mathcal{I}(x) \simeq  1 - \tfrac{1}{1+\kappa x},~x \geq 0, \label{eq:I_approx}
\end{equation}
where $\kappa$ is a large positive number. 
Then, problem \eqref{prob:original_prob} can be approximated by the following continuous problem
\begin{align}
    \min_{\{\mathbf{v}, \mathbf{W}\}} &\left\{\begin{aligned}
        &\sum_{t=1}^T\|\mathbf{W}(t)\|_F^2 + \lambda_1\sum_{t=1}^T\sum_{m=1}^M \left[1 - \tfrac{1}{1+\kappa v_m(t)}\right]\\
        &+\lambda_2\sum_{t=1}^{T-1}\sum_{m=1}^M \left|\tfrac{1}{1+\kappa v_m(t)} - \tfrac{1}{1+\kappa v_m(t+1)}\right|
    \end{aligned}\right\} \label{prob:approximate}\\
    {\rm s.t.}~&~\eqref{cstr:QoS_SOCP},~ \eqref{cstr:power}~{\rm and}~\eqref{cstr:indicator}~{\rm are~satisfied}. \nonumber
\end{align}
Unfortunately, the approximate problem is still difficult due to the non-convex and non-smooth objective. 
As a compromise, we resort to the successive upper-bound minimization (SUM) method \cite{Meisam2013J} to circumvent this difficulty by optimizing a series of approximate upper-bound objective functions.

Specifically, in each iteration, we update $\{\mathbf{v}, \mathbf{W}\}$ by solving the approximate convex problem
\begin{align}
    \min_{\{\mathbf{v}, \mathbf{W}\}}~&\sum_{t=1}^T \|\mathbf{W}(t)\|_F^2 + u(\mathbf{v};\bar{\mathbf{v}}) \label{prob:sum_iteration}\\
    {\rm s.t.} ~~&\eqref{cstr:QoS_SOCP},~ \eqref{cstr:power}~{\rm and}~\eqref{cstr:indicator}~{\rm are~satisfied}, \nonumber
\end{align}
where $u(\mathbf{v};\bar{\mathbf{v}})$ is given in \eqref{eq:upper_bound_objective} at the bottom of next page, with $\bar{v}_m(t)$ and $\bar{v}_m(t+1)$ the iterates of $v_m(t)$ and $v_m(t+1)$ in the previous iteration, respectively; $\bar{\mathbf{v}}$ is the collection of $\bar{v}_m(t)$ for $m = 1, 2, \ldots, M$, $t = 1, 2, \ldots, T$. It is easy to show that \eqref{prob:sum_iteration} is a tight upper bound of \eqref{prob:approximate} since
\setcounter{equation}{10}
\begin{align}
    &\!\!\left[1-\tfrac{1}{1+\kappa v_m(t)}\right] \leq 1 - \tfrac{2}{1+\kappa \bar{v}_m(t)} + \tfrac{1+\kappa v_m(t)}{[1+\kappa \bar{v}_m(t)]^2}, \label{eq:approx3}\\
    &\!\!\left|\tfrac{1}{1+\kappa v_m(t+1)} - \tfrac{1}{1+\kappa v_m(t)}\right|
     = \max\!\left\{\begin{aligned}
        &\!\!\tfrac{1}{1+\kappa v_m(t+1)} - \tfrac{1}{1+\kappa v_m(t)},\!\\
        &\!\!\tfrac{1}{1+\kappa v_m(t)} - \tfrac{1}{1+\kappa v_m(t+1)}\!
        \end{aligned}\right\}\nonumber \\
    &~~~\leq \max\!\left\{\begin{aligned}
        &\!\! \tfrac{1}{1+\kappa v_m(t+1)} - \tfrac{2}{1+\kappa\bar{v}_m(t)} + \tfrac{1+\kappa v_m(t)}{[1+\kappa\bar{v}_m(t)]^2},\\
        &\!\! \tfrac{1}{1+\kappa v_m(t)} - \tfrac{2}{1+\kappa\bar{v}_m(t+1)} + \tfrac{1+\kappa v_m(t+1)}{[1+\kappa\bar{v}_m(t+1)]^2}
    \end{aligned}\right\}\!\! \label{eq:approx4}
\end{align}

The SUM-based approach to problem \eqref{prob:approximate} is summarized as Algorithm 1. Notice that problem \eqref{prob:sum_iteration} is convex, and thus can be globally solved by, e.g., CVX \cite{Grant2014M}.
\begin{table}[!htbp]
  \centering
  \begin{tabular}{cl}
    \toprule
    \multicolumn{2}{l}{{\bf Algorithm 1:} The Offline Approach to Problem \eqref{prob:approximate}}  \\
    \toprule
    1.& Initialize $\mathbf{v}$;\\
    2.& \textbf{\texttt{repeat}}\\
    3.& \quad  $\bar{\mathbf{v}} \leftarrow \mathbf{v}$;\\
    4.& \quad  Solve problem \eqref{prob:sum_iteration} to update $\{\mathbf{v}, \mathbf{W}\}$;\\
    5.& \textbf{\texttt{until}} some stopping criterion is satisfied.\\
    \bottomrule
  \end{tabular}
\end{table}

\begin{remark} \label{lemma:1}
According to Theorem 1 in \cite{Meisam2013J}, we claim that every limit point of the iterates generated by Algorithm 1 is a stationary solution of problem \eqref{prob:approximate}.
\end{remark}


In Algorithm 1, we require that $\{\mathbf{H}(t)\}_{t=1}^T$ are known before solving problem \eqref{prob:approximate}. Hence, Algorithm 1 is actually an offline approach, where we optimize $\{\mathbf{v}(t), \mathbf{W}(t)\}_{t=1}^T$ altogether in time slice $T$ or thereafter. However, this is impractical because $\{\mathbf{v}(t), \mathbf{W}(t)\}$ must be ready in time slice $t$ in order to support real-time data transmission. As a remedy, we further develop an online algorithm that successively outputs $\{\mathbf{v}(t), \mathbf{W}(t)\}$ for $t = 1, 2, \ldots, T$.

\section{Online Algorithm for Problem \eqref{prob:original_prob}} \label{sec:online}
\subsection{Framework of the Online Algorithm}
In the online framework, we optimize $\{\mathbf{v}(t), \mathbf{W}(t)\}$ in time slice $t$. According to \eqref{prob:approximate}, the determination of $\mathbf{v}(t)$ requires the knowledge of $\mathbf{v}(t-1)$ and $\mathbf{v}(t+1)$. Notice that $\mathbf{v}(t+1)$ is unknown currently, which depends on $\mathbf{H}(t+1)$. Nevertheless, it is impossible to obtain $\mathbf{H}(t+1)$ in time slice $t$ because of the causality constraint on CSI. Considering that the distribution of $\mathbf{H}(t+1)$ is usually available a prior, we optimize $\{\mathbf{v}(t), \mathbf{W}(t)\}$ according to the previous admissible users $\mathbf{v}(t-1)$, the actual value of current CSI $\mathbf{H}(t)$, and the distribution of future CSI $\mathbf{H}(t+1)$.

In particular, we construct a new objective function for the problem in time slice $t$, comprised of the actual power cost in time slice $t$, based on the real value of $\mathbf{H}(t)$, and the expected power cost in time slice $(t+1)$, determined by the distribution of $\mathbf{H}(t+1)$, see \eqref{eq:power_cost_t} at the bottom of this page.
\setcounter{equation}{14}

To circumvent the stochastic problem minimizing $p_c(t)$, we replace the expectation by an average of $J$ samples based on SAA \cite{Shapiro2006J}. Specifically, according to the channel distribution, we generate $J$ samples of $\mathbf{H}(t+1)$, denoted by $\hat{\mathbf{H}}(t+1,j)$, $j = 1, 2, \ldots, J$. For each $\hat{\mathbf{H}}(t+1,j)$, we have the corresponding $\hat{\mathbf{v}}(t+1,j)$ and $\hat{\mathbf{W}}(t+1,j)$, from which we can get a sample of the power cost in time slice $(t+1)$. Then, $p_c(t)$ is approximated by $\hat{p}_c(t)$, given in \eqref{eq:power_cost_t_saa} at the bottom of this page. Finally, the problem in time slice $t$ is expressed as
\begin{subequations} \label{prob:online}
\begin{align}
    &\min\limits_{\substack{\{\mathbf{v}(t),\mathbf{W}(t)\} \\ \{\hat{\mathbf{v}}(t+1,j), \hat{\mathbf{W}}(t+1,j)\}_{j = 1}^J}}~\hat{p}_c(t) \nonumber \\
    &\quad {\rm s.t.}~~\eqref{cstr:QoS_SOCP},~ \eqref{cstr:power}~{\rm and}~\eqref{cstr:indicator}~{\rm are~satisfied~for~current}~t, \nonumber \\
    &\qquad~~ \left\{
    \begin{aligned}
            &\!\tfrac{\hat{\mathbf{h}}_{m}^\dag(t+1,j)\hat{\mathbf{w}}_{m}(t+1,j) + \hat{v}_m(t+1,j)}{\sqrt{{\sigma^{2} + \sum_{n\neq m}\left|\hat{\mathbf{h}}_{m}^\dag(t+1,j)\hat{\mathbf{w}}_{n}(t+1,j)\right|^2}}} \geq \sqrt{\gamma}, \\
            &{\Im}\{\hat{\mathbf{h}}^\dag_{m}(t+1,j)\hat{\mathbf{w}}_{m}(t+1,j)\} = 0, \,\forall\,m,j,
         \end{aligned} \right. \label{cstr:QoS_sample} \\
    &\qquad\quad\, \|\hat{\mathbf{W}}(t+1,j)\|_F^2 \leq P,~\forall~j, \label{cstr:trans_power_sample} \\
    &\qquad\quad\, \hat{v}_m(t+1,j) \geq 0,~\forall~m,j. \label{cstr:continuous_indicator_sample}
\end{align}
\end{subequations}
We summarize the online approach as Algorithm 2.
\begin{table}[!htbp]
  \centering
  \begin{tabular}{cl}
    \toprule
    \multicolumn{2}{l}{{\bf Algorithm 2:} The Online Approach for Problem \eqref{prob:approximate} }\\
    \toprule
    1.& Initialize $\mathbf{v}(0)$;\\
    2.& \texttt{\textbf{for}} $t = 1, 2, \ldots, T$, \\
    3.& \quad  Update the current CSI $\mathbf{H}(t)$;\\
    4.& \quad  Generate $J$ samples of future CSI, i.e., $\{\hat{\mathbf{H}}(t+1,j)\}_{j=1}^J$, \\
      & \quad  according to the channel distribution; \\
    5.& \quad  Output $\{\mathbf{v}(t), \mathbf{W}(t)\}$ by solving problem~\eqref{prob:online};\\
    6.& \texttt{\textbf{end}}\\
    \bottomrule
  \end{tabular}
\end{table}

The main complexity of Algorithm 2 lies in solving problem \eqref{prob:online}, which may not be easy due to the non-convex and non-smooth objective, and the high dimension caused by SAA. To handle this, we first apply the SUM method to iteratively solve the problem. To alleviate the complexity, we further devise an ADMM-based algorithm to solve the problem in each iteration efficiently. Finally, a low-complexity and scalable algorithm is designed for problem \eqref{prob:online}.

\subsection{Low-Complexity Algorithm for Problem \eqref{prob:online}}

We define $\breve{\mathbf{H}}(r)$, $\breve{\mathbf{v}}(r)$, $\breve{\mathbf{W}}(r)$, $\breve{\lambda}_0(r)$, $\breve{\lambda}_1(r)$, and $\breve{\lambda}_2(r)$, $r = 1, 2, ..., J+1$, to simplify the notations (see Table \ref{tab:variable}). Further, we put the \emph{available} previous admissible user set in $\breve{\mathbf{v}}(J+2)$, denoted by $\mathbf{v}_p \triangleq \mathbf{v}(t-1)$.
{\renewcommand\arraystretch{1.2}
\begin{table}[!htbp]
  \caption{Summary of the Introduced Variables}	 \label{tab:variable}
  \centering
  \begin{tabular}{|r|c|c|c|c|}
    \hline
    $r = $& $1$ & $2,3, \ldots, J$ & $J + 1$ & $J+2$\\
    \hline
    $\breve{\mathbf{H}}(r)=$ & $\mathbf{H}(t)$ & \multicolumn{2}{|c|}{$\hat{\mathbf{H}}(t+1, r-1)$} & n/a\\
    \hline
    $\breve{\mathbf{v}}(r)=$ & $\mathbf{v}(t)$ & \multicolumn{2}{|c|}{$\hat{\mathbf{v}}(t+1, r-1)$} & $\mathbf{v}_p \triangleq \mathbf{v}(t-1)$ \\
    \hline
    $\breve{\mathbf{W}}(r)=$ & $\mathbf{W}(t)$ & \multicolumn{2}{|c|}{$\hat{\mathbf{W}}(t+1, r-1)$} & n/a\\
    \hline
    $\breve{\lambda}_0(r)= $ & $1 $            & \multicolumn{2}{|c|}{$1/J$}                        & n/a\\
    \hline
    $\breve{\lambda}_1(r)= $ & $\lambda_1$     & \multicolumn{2}{|c|}{$\lambda_1/J$}                & n/a\\
    \hline
    $\breve{\lambda}_2(r)= $ & \multicolumn{2}{|c|}{$\lambda_2/J$} & $\lambda_2$                    & n/a\\
    \hline
  \end{tabular}
\end{table}}

Now, problem~\eqref{prob:online} can be expressed in a unified form,
\begin{subequations} \label{prob:online_uniform}
\begin{align}
    \min\limits_{\{\breve{\mathbf{v}},\breve{\mathbf{W}}\}}~&\,\breve{p}_c(t) \nonumber \\
    {\rm s.t.} ~~&\left\{
    \begin{aligned}
            &\tfrac{\breve{\mathbf{h}}_{m}^\dag(r)\breve{\mathbf{w}}_{m}(r) + \breve{v}_m(r)}{\sqrt{{\sigma^{2} + \sum_{n\neq m}\left|\breve{\mathbf{h}}_{m}^\dag(r)\breve{\mathbf{w}}_{n}(r)\right|^2}}} \geq \sqrt{\gamma}, \\
            &{\Im}\{\breve{\mathbf{h}}^\dag_{m}(r)\breve{\mathbf{w}}_{m}(r)\} = 0, \,\forall\,m,r,
         \end{aligned} \right. \label{cstr:QoS_uniform} \\
    &\,\|\breve{\mathbf{W}}(r)\|_F^2 \leq P,~\forall~r, \label{cstr:trans_power_uniform} \\
    &\,\breve{v}_m(r) \geq 0,~\forall~m,r, \label{cstr:continuous_indicator_uniform}
\end{align}
\end{subequations}
where $\breve{\mathbf{v}}$ and $\breve{\mathbf{W}}$ denote $\{\breve{\mathbf{v}}(r)\}_{r=1}^{J+1}$ and $\{\breve{\mathbf{W}}(r)\}_{r=1}^{J+1}$, respectively; the objective $\breve{p}_c(t)$ is given by
\begin{align}
    \breve{p}_c(t) = \sum_{r=1}^{J+1}\! \left\{
    \begin{aligned}
    &\!\breve{\lambda}_0(r)\|\breve{\mathbf{W}}(r)\|_F^2 + \breve{\lambda}_1(r)\sum_{m=1}^M \! \left[1 - \tfrac{1}{1+\kappa\breve{v}_m(r)}\right] \\
    &\!\!+\breve{\lambda}_2(r)\sum_{m=1}^M \! \left|\tfrac{1}{1+\kappa\breve{v}_m(r+1)} - \tfrac{1}{1+\kappa\breve{v}_m(1)}\right|
    \end{aligned}\!\right\} \nonumber
\end{align}

Again, to handle the non-convex and non-smooth objective, we apply the SUM method and solve problem \eqref{prob:online_uniform} iteratively. In each iteration, the problem with an approximate objective{\footnote{Originally, the objective of \eqref{prob:online_uniform_iteration} is a upper bound of $\breve{p}_c(t)$, similar as \eqref{eq:upper_bound_objective}. Here we remove some constant terms in the upper-bound function to simplify the formulation.}} is given by
\begin{subequations} \label{prob:online_uniform_iteration}
\begin{align}
    &\min_{\{\breve{\mathbf{v}}, \breve{\mathbf{W}}, \mathbf{a}\}} ~\sum_{r=1}^{J + 1} \! \left\{
    \begin{aligned}
    & \breve{\lambda}_0(r)\|\breve{\mathbf{W}}(r)\|_F^2 \\
    &\! + \sum_{m=1}^M \!\! \left[\breve{\lambda}_1(r)\tfrac{\kappa\breve{v}_m(r)}{[1+\kappa\tilde{v}_m(r)]^2} + \breve{\lambda}_2(r)a_m(r)\right] \!
    \end{aligned}\right\} \nonumber \\
    &\quad\,{\rm s.t.}~~ \eqref{cstr:QoS_uniform},~ \eqref{cstr:trans_power_uniform},~{\rm and}~ \eqref{cstr:continuous_indicator_uniform}~{\rm are~satisfied},\nonumber \\
    &\qquad~~\begin{cases}
    \!a_m(r) \geq \tfrac{1}{1+\kappa\breve{v}_m(1)} - \tfrac{2}{1+\kappa\tilde{v}_m(r+1)} + \tfrac{1+\kappa\breve{v}_m(r+1)}{[1+\kappa\tilde{v}_m(r+1)]^2},\\
    \!a_m(r) \geq \tfrac{1}{1+\kappa\breve{v}_m(r+1)} - \tfrac{2}{1+\kappa\tilde{v}_m(1)} + \tfrac{1+\kappa\breve{v}_m(1)}{[1+\kappa\tilde{v}_m(1)]^2},
    \end{cases} \nonumber\\
    &\qquad\qquad\qquad\qquad\qquad \forall~m,~{\rm and}~r = 1, 2, \ldots, J, \label{cstr:switch_onlin_uniform_iteration_a}\\
    &\qquad~~\begin{cases}
    a_m(r) \geq \tfrac{1}{1+\kappa \breve{v}_m(1)} - \!\left[\tfrac{1}{1+\kappa {v}_{p,m}}\right]_q\!, \\
    a_m(r) \geq \left[\!\tfrac{1}{1+\kappa v_{p,m}}\!\right]_q - \tfrac{2}{1+\kappa\tilde{v}_m(1)} + \tfrac{1+\kappa\breve{v}_m(1)}{[1+\kappa\tilde{v}_m(1)]^2},
    \end{cases} \nonumber\\
    &\qquad\qquad\qquad\qquad\qquad \forall~m,~{\rm and}~r = J+1,  \label{cstr:switch_onlin_uniform_iteration_b}
\end{align}
\end{subequations}
where $\mathbf{a}$ is the collection of $a_m(r)$, for $m = 1, 2,\ldots, M$, and $r = 1, 2, \ldots, J+1$; $\tilde{v}_m(r)$ and $\tilde{v}_m(r+1)$ are the iterates of $\breve{v}_m(r)$ and $\breve{v}_m(r+1)$ in the previous iteration; $[\cdot]_q$ quantizes the argument to the nearest integer (we get 0 or 1 here); $v_{p,m}$ is the $m$th element of $\mathbf{v}_p$, i.e., $v_{p,m} = v_m(t-1)$.
\begin{figure*}[b]
  \hrulefill
  \setcounter{mytempeqncnt}{\value{equation}}
  \setcounter{equation}{19}
  \begin{align}
    &\mathcal{L}_{\rho}\!\left(\substack{\breve{\mathbf{v}}, \breve{\mathbf{W}}, \mathbf{a}, \mathbf{b}, \mathbf{c}, \mathbf{E}, \mathbf{x}, \mathbf{y}, \mathbf{z}, \mathbf{s},\\
    \boldsymbol{\Omega}, \boldsymbol{\theta}, \boldsymbol{\phi}, \boldsymbol{\epsilon}, \boldsymbol{\delta}, \boldsymbol{\tau}, \boldsymbol{\eta}}\right)  = \sum\limits_{r=1}^{J+1}\left\{\breve{\lambda}_0(r)\|\breve{\mathbf{W}}(r)\|_F^2 + \sum\limits_{m=1}^M \! \left[ \breve{\lambda}_1(r)\tfrac{\kappa \breve{v}_m(r)}{\left[1+\kappa\tilde{v}_m(r)\right]^2}  + \breve{\lambda}_2(r)a_m(r)\right]\right\} \nonumber \\
    &\qquad + \sum\limits_{r=1}^{J+1}\left\{{\Re}\!\left\{{\rm Tr}\!\left[\boldsymbol{\Omega}^\dag(r)\!\left(\mathbf{E}(r) - [\breve{\mathbf{H}}^\dag(r)\breve{\mathbf{W}}(r), \sigma\mathbf{1}]\right)\right]\right\} + \frac{\rho}{2}\left\|\mathbf{E}(r) - [\breve{\mathbf{H}}^\dag(r)\breve{\mathbf{W}}(r), \sigma\mathbf{1}]\right\|_F^2\right\} \nonumber \\
    &\qquad + \sum\limits_{r = 1}^{J+1}\sum\limits_{m=1}^M \left\{\theta_m(r)\left[a_m(r)-b_m(r)\right] + \frac{\rho}{2}\left[a_m(r)-b_m(r)\right]^2 + \phi_m(r)\left[a_m(r)-c_m(r)\right] + \frac{\rho}{2}\left[a_m(r)-c_m(r)\right]^2 \right\} \nonumber \\
    &\qquad + \sum\limits_{r = 1}^{J+1}\sum\limits_{m=1}^M \left\{\epsilon_m(r)\left[x_m(r)-1 - \kappa \breve{v}_m(1)\right] + \frac{\rho}{2}\left[x_m(r)-1 - \kappa \breve{v}_m(1)\right]^2\right\} \label{eq:arg_lag}\\
    &\qquad + \sum\limits_{r = 1}^{J+1}\sum\limits_{m=1}^M \left\{\delta_m(r)\!\left[y_m(r)-\tfrac{1+\kappa \breve{v}_m(1)}{\left[1+\kappa \tilde{v}_m(1)\right]^2}\right] + \frac{\rho}{2}\!\left[y_m(r)-\tfrac{1+\kappa \breve{v}_m(1)}{\left[1+\kappa \tilde{v}_m(1)\right]^2}\right]^2\right\} \nonumber \\
    &\qquad + \sum\limits_{r = 1}^{J}\sum\limits_{m=1}^M \left\{\tau_m(r) \left[z_m(r)-1 - \kappa \breve{v}_m(r+1)\right] + \frac{\rho}{2}\left[z_m(r)-1 - \kappa \breve{v}_m(r+1)\right]^2\right\} \nonumber \\
    &\qquad + \sum\limits_{t = 1}^{J}\sum\limits_{m=1}^M \left\{\eta_m(r)\! \left[s_m(r)-\tfrac{1+\kappa \breve{v}_m(r+1)}{\left[1+\kappa \tilde{v}_m(r+1)\right]^2}\right] + \frac{\rho}{2}\!\left[s_m(r)-\tfrac{1+\kappa \breve{v}_m(r+1)}{\left[1+\kappa \tilde{v}_m(r+1)\right]^2}\right]^2\right\}, \nonumber
  \end{align}
  \setcounter{equation}{\value{mytempeqncnt}}
  \hrulefill
  \vspace{6pt}
  \setcounter{equation}{20}
  \begin{subequations}\label{eq:admm_iteration}
    \begin{align}
    &\left\{\breve{\mathbf{W}}^{k+1}, \mathbf{b}^{k+1}, \mathbf{c}^{k+1}, \mathbf{x}^{k+1}, \mathbf{y}^{k+1}, \mathbf{z}^{k+1}, \mathbf{s}^{k+1}\right\} \leftarrow \mathop{\arg\min}_{\{\breve{\mathbf{W}}, \mathbf{b}, \mathbf{c}, \mathbf{x}, \mathbf{y}, \mathbf{z}, \mathbf{s}\}} \mathcal{L}_{\rho}\!\left(\substack{\breve{\mathbf{v}}^k, \breve{\mathbf{W}}, \mathbf{a}^k, \mathbf{b}, \mathbf{c}, \mathbf{E}^k, \mathbf{x}, \mathbf{y}, \mathbf{z}, \mathbf{s},\\
    \boldsymbol{\Omega}^k, \boldsymbol{\theta}^k, \boldsymbol{\phi}^k, \boldsymbol{\epsilon}^k, \boldsymbol{\delta}^k, \boldsymbol{\tau}^k, \boldsymbol{\eta}^k}\right) \label{eq:admm_iteration_subprob1}\\
    &\qquad\qquad\qquad\qquad\qquad\qquad\qquad\qquad\qquad\qquad\qquad\quad~ {\rm s.t.}~~\eqref{cstr:trans_power_uniform}, \eqref{cstr:admm_xadycb_1}-\eqref{cstr:admm_ac}~{\rm are~satisfied}, \nonumber \\
    &\left\{\breve{\mathbf{v}}^{k+1}, \mathbf{a}^{k+1}, \mathbf{E}^{k+1}\right\} \leftarrow \mathop{\arg\min}_{\{\breve{\mathbf{v}}, \mathbf{a}, \mathbf{E}\}} \mathcal{L}_{\rho}\!\left(\substack{\breve{\mathbf{v}}, \breve{\mathbf{W}}^{k+1}, \mathbf{a}, \mathbf{b}^{k+1}, \mathbf{c}^{k+1}, \mathbf{E}, \mathbf{x}^{k+1}, \mathbf{y}^{k+1}, \mathbf{z}^{k+1}, \mathbf{s}^{k+1},\\
    \boldsymbol{\Omega}^k, \boldsymbol{\theta}^k, \boldsymbol{\phi}^k, \boldsymbol{\epsilon}^k, \boldsymbol{\delta}^k, \boldsymbol{\tau}^k, \boldsymbol{\eta}^k}\right)\label{eq:admm_iteration_subprob2}\\
    &\qquad\qquad\qquad\qquad\qquad\quad {\rm s.t.}~~  \eqref{cstr:admm_QoS}~{\rm is~satisfied}, \nonumber \\
    &\begin{cases}
        \boldsymbol{\Omega}^{k+1}(r) \leftarrow \boldsymbol{\Omega}^{k}(r) + \rho \left( \mathbf{E}^{k+1}(r) - [\breve{\mathbf{H}}^\dag(r)\breve{\mathbf{W}}^{k+1}(r), \sigma\mathbf{1}]\right),\\
        \theta_m^{k+1}(r) \leftarrow \theta_m^{k}(r) + \rho [a_m^{k+1}(r) - b_m^{k+1}(r)],\qquad\qquad~~~\, \phi_m^{k+1}(r) \leftarrow \phi_m^{k}(r) + \rho [a_m^{k+1}(r) - c_m^{k+1}(r)],\\
        \epsilon_m^{k+1}(r) \leftarrow \epsilon_m^{k}(r) + \rho [x_m^{k+1}(r) - 1 - \kappa \breve{v}_m^{k+1}(1)],~~~~~~~\,\; \delta_m^{k+1}(r) \leftarrow \delta_m^{k}(r) + \rho \left(y_m^{k+1}(r) - \tfrac{1+\kappa\breve{v}^{k+1}_m(1)}{[1+\kappa\tilde{v}_m(1)]^2}\right),\\
        \tau_m^{k+1}(r) \leftarrow \tau_m^{k}(r) + \rho [z_m^{k+1}(r) - 1 - \kappa\breve{v}^{k+1}(r+1)],~~~ \eta_m^{k+1}(r) \leftarrow \eta_m^{k}(r) + \rho \left(s_m^{k+1}(r) - \tfrac{1+\kappa\breve{v}^{k+1}_m(r+1)}{[1+\kappa\tilde{v}_m(r+1)]^2}\right),\\
    \end{cases} \label{eq:admm_iteration_lagrangian}
    \end{align}
  \end{subequations}
\end{figure*}
\setcounter{equation}{17}

Because of the high dimension caused by SAA, people may prefer a low-complexity approach to problem \eqref{prob:online_uniform_iteration}. Motivated by this, we further design an efficient ADMM-based algorithm for it. To this end, we introduce some auxiliary variables:
\begin{subequations}
\begin{align}
    &\mathbf{E}(r) = [\breve{\mathbf{H}}^\dag(r)\breve{\mathbf{W}}(r), \sigma\mathbf{1}], ~~\forall~r, \label{cstr:admm_f}\\
    &b_m(r) = a_m(r),~c_m(r) = a_m(r),~~ \forall\,m,r, \label{cstr:admm_xy}\\
    &x_m(r) = 1+\kappa \breve{v}_m(1),~y_m(r) = \tfrac{1+\kappa \breve{v}_m(1)}{\left[1+\kappa \tilde{v}_m(1)\right]^2},\, \forall\,m,r, \label{cstr:admm_ab}\\
    &z_m(r) = 1+\kappa \breve{v}_m(r+1),~s_m(r) = \tfrac{1+\kappa \breve{v}_m(r+1)}{\left[1+\kappa \tilde{v}_m(r+1)\right]^2}, \nonumber \\
    &\qquad\qquad\qquad\qquad\quad~~\,\forall~m,~{\rm and}~r = 1, 2, \ldots, J. \label{cstr:admm_cd}
\end{align}
\end{subequations}
Then, problem \eqref{prob:online_uniform_iteration} can be equivalently expressed as
\begin{subequations} \label{prob:online_decouple}
\begin{align}
    \!\!\min_{\left\{\substack{\breve{\mathbf{v}},\breve{\mathbf{W}}, \mathbf{a},\mathbf{b},\mathbf{c} \\ \mathbf{E},\mathbf{x},\mathbf{y},\mathbf{z},\mathbf{s}}\right\}}&\, \sum_{r=1}^{J + 1} \! \left\{
    \begin{aligned}
    & \breve{\lambda}_0(r)\|\breve{\mathbf{W}}(r)\|_F^2 \\
    &\!\! + \sum_{m=1}^M \!\! \left[\breve{\lambda}_1(r)\tfrac{\kappa\breve{v}_m(r)}{[1+\kappa\tilde{v}_m(r)]^2} + \breve{\lambda}_2(r)a_m(r)\right] \!
    \end{aligned}\right\} \nonumber \\
    {\rm s.t.}~~~~ &\left\{\begin{aligned}
    &\!e_m^m(r) + \breve{v}_m(r) \geq \sqrt{\gamma}\|\mathbf{e}_{-m}^m(r)\|_2,\\
    &\!{\Im}\{e_m^m(r)\} = 0,~~~\forall~m,r,
    \end{aligned}\right. \label{cstr:admm_QoS}\\
    &\begin{cases}
    b_m(r) \geq \tfrac{1}{x_m(r)} - \tfrac{2}{1+\kappa\tilde{v}_m(r+1)} + s_m(r),\\
    c_m(r) \geq \tfrac{1}{z_m(r)} - \tfrac{2}{1+\kappa\tilde{v}_m(1)} + y_m(r),
    \end{cases}\nonumber\\
    &\qquad\qquad\quad~~ \forall~m,\,{\rm and}~ r = 1, 2, \ldots, J, \label{cstr:admm_xadycb_1}\\
    &\begin{cases}
    b_m(r) \geq \tfrac{1}{x_m(r)} - \!\left[\tfrac{1}{1+\kappa v_{p,m}}\right]_q,\\
    c_m(r) \geq \left[\tfrac{1}{1+\kappa v_{p,m}}\right]_q \! - \tfrac{2}{1+\kappa\tilde{v}_m(1)} + y_m(r),
    \end{cases}\nonumber\\
    &\qquad\qquad\quad~~ \forall~m,~{\rm and}~r = J+1, \label{cstr:admm_xadycd_2} \\
    &~x_m(r) \geq 1,~z_m(r) \geq 1, ~\forall~m,r, \label{cstr:admm_ac} \\
    &~\eqref{cstr:trans_power_uniform}, \eqref{cstr:admm_f} - \eqref{cstr:admm_cd}~{\rm are~satisfied}. \nonumber
\end{align}
\end{subequations}
From \eqref{cstr:admm_f}, we know that \eqref{cstr:admm_QoS} is actually the QoS constraint \eqref{cstr:QoS_uniform}, where $\mathbf{e}^m(r) = [\breve{\mathbf{h}}_m^\dag(r)\breve{\mathbf{W}}(r), \sigma] \in \mathbb{C}^{1 \times (M+1)}$ is the $m$th row vector of $\mathbf{E}(r) \in \mathbb{C}^{M\times (M+1)}$; $e^m_m(r) = \breve{\mathbf{h}}_m^\dag(r)\breve{\mathbf{w}}_m(r)$ is the $m$th element of $\mathbf{e}^m(r)$, which is also the $m$th diagonal element of $\mathbf{E}(r)$, $m = 1, 2, \ldots, M$; $\mathbf{e}^m_{-m}(r) \in \mathbb{C}^{1\times M}$ is obtained by removing $e^m_m(r)$ from $\mathbf{e}^m(r)$, i.e., $\mathbf{e}_{-m}^m(r) \triangleq [e^m_1(r),$ $\ldots, e^m_{m-1}(r), e^m_{m+1}(r), \ldots, e^m_{M+1}(r)]$. From \eqref{cstr:admm_ab} and \eqref{cstr:admm_cd}, \eqref{cstr:continuous_indicator_uniform} is equivalently addressed by \eqref{cstr:admm_ac}.

In problem~\eqref{prob:online_decouple}, the equality constraints, i.e., \eqref{cstr:admm_f} -- \eqref{cstr:admm_cd}, can be moved to the objective by the augmented Lagrangian method \cite{Hestenses1969J}, which generates the objective (also referred to as the partial augmented Lagrangian function) in \eqref{eq:arg_lag}, where $\{\boldsymbol{\Omega}, \boldsymbol{\theta}, \boldsymbol{\phi}, \boldsymbol{\epsilon}, \boldsymbol{\delta}, \boldsymbol{\tau}, \boldsymbol{\eta}\}$ are the Lagrangian multipliers. Following the ADMM framework \cite{Boyd2011J}, we divide $\{\breve{\mathbf{v}}, \breve{\mathbf{W}},\mathbf{a}, \mathbf{b}, \mathbf{c}, \mathbf{E}, \mathbf{x}, \mathbf{y},$ $\mathbf{z}, \mathbf{s}\}$ into $\{\breve{\mathbf{W}}, \mathbf{b}, \mathbf{c}, \mathbf{x}, \mathbf{y}, \mathbf{z}, \mathbf{s}\}$ and $\{\breve{\mathbf{v}}, \mathbf{a}, \mathbf{E}\}$, and then solve problem \eqref{prob:online_decouple} by alternating among the steps in \eqref{eq:admm_iteration}, where $k$ is the iteration index. In particular, \eqref{eq:admm_iteration_subprob1} and \eqref{eq:admm_iteration_subprob2} can be divided into several low-dimensional subproblems and solved efficiently, which thereby effectively alleviates the complexity arising from high dimension. In the rest of this subsection, we show the solution of each subproblem concretely. To simplify the notation, the iteration index $k$ is omitted when there is no ambiguity.
\setcounter{equation}{21}

\subsubsection{Updating $\{\breve{\mathbf{W}}, \mathbf{b}, \mathbf{c}, \mathbf{x}, \mathbf{y}, \mathbf{z}, \mathbf{s}\}$}
It is easy to observe that problem \eqref{eq:admm_iteration_subprob1} can be divided into $(J+1)$ problems of $\breve{\mathbf{W}}(r)$, $(J+1)M$ problems of $\{b_m(r), x_m(r), s_m(r)\}$, and $(J+1)M$ problems of $\{c_m(r), y_m(r), z_m(r)\}$, $\forall~m,r$. We update them individually.

\textcircled{\small 1}~Let $\mathbf{E}_{w}(r)$ and $\boldsymbol{\Omega}_w(r)$ be the left $M \times M$ sub-matrices of $\mathbf{E}(r)$ and $\boldsymbol{\Omega}(r) \in \mathbb{C}^{M \times (M+1)}$, respectively. The subproblem with respect to $\breve{\mathbf{W}}(r)$ is expressed as
\begin{align}
    \min_{\breve{\mathbf{W}}(r)}\, &\left\{\begin{aligned}
    &\breve{\lambda}_0(r)\|\breve{\mathbf{W}}(r)\|_F^2 +\frac{\rho}{2}\|\mathbf{E}_w(r) - \breve{\mathbf{H}}^\dag(r)\breve{\mathbf{W}}(r)\|_F^2 \\
    &- {\Re}\left\{{\rm Tr}\!\left[\boldsymbol{\Omega}_w^\dag(r)\breve{\mathbf{H}}^\dag(r)\breve{\mathbf{W}}(r)\right]\right\}
    \end{aligned}\right\} \nonumber \\
    {\rm s.t.}~ &~ \|\breve{\mathbf{W}}(r)\|_F^2 \leq P, \label{prob:admm_update_w}
\end{align}
and can be easily solved as
\begin{align}
    \breve{\mathbf{W}}(r) = &\left[\rho \breve{\mathbf{H}}(r)\breve{\mathbf{H}}^\dag(r) + 2[\breve{\lambda}_0(r)+\alpha(r)]\mathbf{I}\right]^{-1} \nonumber \\
    &\qquad\qquad\quad \times \breve{\mathbf{H}}(r)[\rho\mathbf{E}_w(r)+\boldsymbol{\Omega}_w(r)], \label{eq:admm_update_w}
\end{align}
where $\alpha(r)$ is the Lagrangian multiplier with $\|\breve{\mathbf{W}}(r)\|_F^2 \leq P$. According to the Karush-Kuhn-Tucker (KKT) conditions \cite{Boyd2004M}, $\alpha(r)$ can be easily determined by bisection.

\textcircled{\small 2}~The subproblem with respect to $\{b_m(r), x_m(r), s_m(r)\}$ is expressed as{\footnote{Problem \eqref{prob:admm_update_xad} works for $r = 1, 2, \ldots, J$. In the case of $r = J+1$, there is no $s_m(r)$ and we get a simplified version of \eqref{prob:admm_update_xad}. Due to the space limitation, we omit the details of solving the problem with $r = J+1$.}}
\begin{subequations}\label{prob:admm_update_xad}
\begin{align}
    \min_{\left\{\substack{b_m(r)\\ x_m(r)\\ s_m(r)}\right\}} &~\frac{\rho}{2}\!\left\{\begin{aligned}
    &\!\! \left[b_m(r) - a_m(r) - \tfrac{\theta_m(r)}{\rho}\right]^2\\
    &\! +\!\left[x_m(r)-1 - \kappa \breve{v}_m(1) + \tfrac{\epsilon_m(r)}{\rho}\right]^2 \\
    &\! +\! \left[s_m(r)-\tfrac{1+\kappa \breve{v}_m(r+1)}{\left[1+\kappa \tilde{v}_m(r+1)\right]^2} + \tfrac{\eta_m(r)}{\rho}\right]^2
    \end{aligned}\right\} \nonumber \\
    {\rm s.t.}~~\,  &\, b_m(r) \geq \tfrac{1}{x_m(r)} - \tfrac{2}{1+\kappa\tilde{v}_m(r+1)} + s_m(r), \label{cstr:xad} \\
    &\, x_m(r) \geq 1.\label{cstr:a}
\end{align}
\end{subequations}
Based on the first-order optimality condition, we solve $\{b_m(r),$ $x_m(r),$ $s_m(r)\}$ as
\begin{subequations} \label{eq:KKT_xad}
\begin{align}
        & b_m(r) = a_m(r) + \tfrac{\theta_m(r)}{\rho} + \tfrac{\beta_m(r)}{\rho}, \label{eq:KKT_xad_b}\\
        & x_m(r) = \left[{\rm root} \! \left\{\substack{x_m(r) - 1 - \kappa \breve{v}_m(1) + \frac{\epsilon_m(r)}{\rho} = \frac{\beta_m(r)}{\rho x_m^2(r)}}\right\}\right]_1^+\!,  \label{eq:KKT_xad_x}\\
        & s_m(r) = \tfrac{1+\kappa \breve{v}_m(r+1)}{\left[1 + \kappa \tilde{v}_m(r+1)\right]^2} - \tfrac{\eta_m(r)}{\rho} - \tfrac{\beta_m(r)}{\rho},\label{eq:KKT_xad_s}
\end{align}
\end{subequations}
where $\beta_m(r)$ is the Lagrangian multiplier; $[\cdot]_1^+ \triangleq \max\{1,\cdot\}$; ${\rm root}\{\cdot\}$ returns the value of $x_m(r)$ satisfying the cubic equation inside.

Notice that to determine $\{b_m(r), x_m(r), s_m(r)\}$, we do not need to really solve the cubic equation. In the case that \eqref{cstr:xad} is satisfied at $\beta_m(r) = 0$, i.e.,
\begin{align}
    \Gamma_m(r) &\triangleq a_m(r) + \tfrac{\theta_m(r)+\eta_m(r)}{\rho} + \tfrac{1 + \kappa\left[2\tilde{v}_m(r+1) - \breve{v}_m(r+1)\right]}{\left[1+\kappa \tilde{v}_m(r+1)\right]^2} \nonumber \\
    & \geq \tfrac{1}{\left[1 + \kappa \breve{v}_m(1)-\frac{\epsilon_m(r)}{\rho}\right]_1^+},
\end{align}
we have $\beta_m(r) = 0$ in \eqref{eq:KKT_xad}. Otherwise, we find some $\beta_m(r) > 0$ such that \eqref{cstr:xad} holds for equality. Then, we must have
\begin{align}
    \tfrac{1}{x_m(r)} &= b_m(r) + \tfrac{2}{1+\kappa\tilde{v}_m(r+1)} - s_m(r) \nonumber\\
                      &= \tfrac{2\beta_m(r)}{\rho}  + \Gamma_m(r).\label{eq:am_inv}
\end{align}
After that, we insert \eqref{eq:am_inv} into \eqref{eq:KKT_xad_x} and then get
\begin{equation}
    x_m(r) = \left[\substack{1 + \kappa \breve{v}_m(1) - \frac{\epsilon_m(r)}{\rho} + \frac{\beta_m(r)[2\beta_m(r) + \rho\Gamma_m(r)]^2}{\rho^3}}\right]_1^+ \label{eq:am}
\end{equation}
According to \eqref{eq:am_inv} and \eqref{eq:am}, we have
\begin{equation}
    \left[\substack{1 + \kappa \breve{v}_m(1) - \frac{\epsilon_m(r)}{\rho} + \frac{\beta_m(r)[2\beta_m(r) + \rho\Gamma_m(r)]^2}{\rho^3}}\right]_1^+ = \tfrac{\rho}{2\beta_m(r) + \rho\Gamma_m(r)}. \label{eq:beta}
\end{equation}
Then, $\beta_m(r)$ can be solved by performing bisection search. In addition, the range of bisection can be identified as
\begin{equation}
    \beta_m(r) \in \left[\max\!\left\{0, \tfrac{-\rho\Gamma_m(r)}{2}\right\}\!,~ \tfrac{\rho - \rho\Gamma_m(r)}{2}\right].
\end{equation}

\textcircled{\small 3}~$\{c_m(r), y_m(r), z_m(r)\}$ can be updated similarly.

\subsubsection{Updating $\{\breve{\mathbf{v}}, \mathbf{a}, \mathbf{E}\}$}
Problem \eqref{eq:admm_iteration_subprob2} can be divided into $(J+1)M$ problems of $\{\mathbf{e}^{m}(r),\breve{v}_m(r)\}$ and $(J+1)M$ problems of $a_m(r)$, $\forall\,m,r$. They can also be solved individually.

\textcircled{\small 1}~Define $\mathbf{G}(r) \triangleq \rho[\breve{\mathbf{H}}^\dag(r)\breve{\mathbf{W}}(r),\sigma\mathbf{1}] - \boldsymbol{\Omega}(r)$. Similar as $\mathbf{e}^m(r)$, let $\mathbf{g}^m(r) \in \mathbb{C}^{(M+1) \times 1}$ denote the $m$th row of $\mathbf{G}(r)$. The subproblem of $\{\mathbf{e}^m(r), \breve{v}_m(r)\}$ is expressed as
\begin{subequations}\label{prob:admm_update_fs}
\begin{align}
    \min_{\left\{\substack{\mathbf{e}^m(r)\\ \breve{v}_m(r)}\right\}}&\!\left\{\begin{aligned}
    &\!\tfrac{\breve{\lambda}_1(r)\kappa \breve{v}_m(r)}{\left[1+\kappa\tilde{v}_m(r)\right]^2}+ \tfrac{\rho}{2}\|\mathbf{e}^m(r) - \tfrac{\mathbf{g}^m(r)}{\rho}\|_2^2 \\
    &\!\!+ \tfrac{\iota(r)\rho}{2}\sum\nolimits_{j=1}^{J+1}\!\left[\kappa \breve{v}_m(r) + 1 - x_m(j) - \tfrac{\epsilon_m(j)}{\rho} \right]^2\\
    &\!\!+ \tfrac{\iota(r)\rho}{2}\sum\nolimits_{j=1}^{J+1}\!\left[\tfrac{\kappa \breve{v}_m(r) + 1}{\left[1+\kappa \tilde{v}_m(r)\right]^2} - y_m(j)- \tfrac{\delta_m(j)}{\rho}\right]^2 \\
    &\!\!+ \tfrac{[1 - \iota(r)]\rho}{2}\!\left[\kappa \breve{v}_m(r) + 1 - z_m(r-1) - \tfrac{\tau_m(r-1)}{\rho}\right]^2 \!\\
    &\!\!+ \tfrac{[1 - \iota(r)]\rho}{2}\!\left[\tfrac{\kappa \breve{v}_m(r) + 1}{\left[1+\kappa \tilde{v}_m(r)\right]^2} - s_m(r-1) - \tfrac{\eta_m(r-1)}{\rho}\right]^2
    \end{aligned}\right\} \nonumber \\
    {\rm s.t.}~~ &\! \begin{cases}
                    e^m_m(r) + \breve{v}_m(r) \geq \sqrt{\gamma}\|\mathbf{e}^m_{-m}(r)\|_2,\\
                    {\Im}\{e^m_m(r)\} = 0,
                 \end{cases}\\
                 &\iota(r) \triangleq \begin{cases}
                    1, ~~r = 1,\\
                    0, ~~r = 2,3,\ldots, J+1,
                 \end{cases}
\end{align}
\end{subequations}
According to the first-order optimality conditions, we get
\begin{subequations} \label{eq:solution_f}
\begin{align}
    &\breve{v}_m(r) =\tfrac{f_m(r) + \mu_m(r)}{\rho \cdot q_m(r)},\\
    &e^m_m(r) = \tfrac{{\Re}\{g_m^m(r)\}+\mu_m(r)}{\rho},\\
    &\mathbf{g}_{-m}^m(r) -\rho \mathbf{e}^m_{-m}(r) \in \mu_m(r)\sqrt{\gamma}\partial\|\mathbf{e}^m_{-m}(r)\|_2,
\end{align}
\end{subequations}
where $\mu_m(r)$ is the Lagrangian multiplier; $g_m^m(r)$ is the $m$th element of $\mathbf{g}^m(r)$, and $\mathbf{g}_{-m}^m(r)$ is obtained by removing $g_m^m(r)$ from $\mathbf{g}^m(r)$; $f_m(r)$ and $q_m(r)$ are respectively defined as \eqref{eq:chi} at the top of the next page and \eqref{eq:xi} below, \setcounter{equation}{32}
\begin{equation}
    q_m(r) \triangleq \kappa^2[J\cdot\iota(r) + 1]\left\{1 + \tfrac{1}{\left[1+\kappa \tilde{v}_m(r)\right]^4}\right\}. \label{eq:xi}
\end{equation}
In addition, $\partial\|\cdot\|_2$ is the subgradient \cite{Hestenses1969J} of the non-smooth $\ell_2$-norm function $\|\cdot\|_2$, defined as
\begin{equation} \label{eq:subgradient}
    \partial\|\boldsymbol{\zeta}\|_2 \triangleq \left\{
    \begin{aligned}
        &\tfrac{\boldsymbol{\zeta}}{\|\boldsymbol{\zeta}\|_2},~ \boldsymbol{\zeta} \neq \mathbf{0}, \\
        &\{\boldsymbol{\xi}\;|~ \|\boldsymbol{\xi}\|_2 \leq 1\}, ~ \boldsymbol{\zeta} = \mathbf{0},
    \end{aligned}\right.
\end{equation}
where $\boldsymbol{\xi}$ is an arbitrary vector having the same dimension with $\boldsymbol{\zeta}$. Utilizing the method in \cite{Lin2014C}, we combine \eqref{eq:solution_f}, \eqref{eq:subgradient} and the KKT conditions, and then solve $\{\mathbf{e}^m(r), \breve{v}_m(r)\}$ as follows
\begin{align}
    &\texttt{if}~\|\mathbf{g}_{-m}^m(r)\|_2 \leq \tfrac{\sqrt{\gamma} \cdot \left[-f_m(r)- q_m(r){\Re}\{g_m^m(r)\}\right]_0^+}{1+q_m(r)} \nonumber \\
    &\qquad \breve{v}_m(r) = \tfrac{f_m(r) + \mu_m(r)}{\rho \cdot q_m(r)}, \nonumber\\
    &\qquad e^m_m(r) = \tfrac{{\Re}\{g_m^m(r)\}+\mu_m(r)}{\rho}, \nonumber\\
    &\qquad \mathbf{e}^m_{-m}(r) = \mathbf{0}, \nonumber\\
    &\qquad \mu_m(r) = \tfrac{\left[-f_m(r)- q_m(r){\Re}\{g_m^m(r)\}\right]_0^+}{1+q_m(r)}, \nonumber \\
    &\texttt{else},\label{eq:admm_update_fs}\\
    &\qquad \breve{v}_m(r) = \tfrac{f_m(r) + \mu_m(r)}{\rho\cdot q_m(r)}, \nonumber\\
    &\qquad e^m_m(r) = \tfrac{{\Re}\{g_m^m(r)\}+\mu_m(r)}{\rho},\nonumber \\
    &\qquad \mathbf{e}^m_{-m}(r) = \tfrac{\|\mathbf{g}_{-m}^m(r)\|_2 - \mu_m(r)\sqrt{\gamma}}{\rho} \cdot\tfrac{\mathbf{g}_{-m}^m(r)}{\|\mathbf{g}_{-m}^m(r)\|_2}, \nonumber \\
    &\qquad \mu_m(r) = \tfrac{\left[q_m(r)\left(\sqrt{\gamma}\|\mathbf{g}_{-m}^m(r)\|_2 - {\Re}\{g_m^m(r)\}\right)-f_m(r)\right]_0^+}{1+(1+\gamma)q_m(r)}. \nonumber
\end{align}
where $[\cdot]_0^+ \triangleq \max\{0,\cdot\}$.
\begin{figure*}[t]
  \setcounter{mytempeqncnt}{\value{equation}}
  \setcounter{equation}{31}
  \begin{align}
    &f_m(r) \triangleq \kappa\rho\left\{\begin{aligned}
    &\!\tfrac{\iota(r)\sum_{j=1}^{J+1}\epsilon_m(j) + [1 - \iota(r)]\tau_m(r-1)}{\rho} + \tfrac{\iota(r)\sum_{j=1}^{J+1}\delta_m(j)+ [1-\iota(r)]\eta_m(r-1) - \breve{\lambda}_1(r)}{\rho\left[1+\kappa\tilde{v}_m(r)\right]^2}  +\iota(r)\sum\nolimits_{j=1}^{J+1}[x_m(j) - 1] \\
    & + [1 - \iota(r)][z_m(r-1) - 1] +\tfrac{\iota(r)\sum_{j=1}^{J+1}\left(y_m(j)-\frac{1}{\left[1+\kappa\tilde{v}_m(r)\right]^2}\right)+ [1 - \iota(r)]\left(s_m(r-1) - \frac{1}{\left[1+\kappa\tilde{v}_m(r)\right]^2}\right)}{\left[1+\kappa \tilde{v}_m(r)\right]^2}
    \end{aligned}\right\}, \label{eq:chi}
  \end{align}
  \setcounter{equation}{\value{mytempeqncnt}}
  \hrulefill
\end{figure*}
\setcounter{equation}{35}

\textcircled{\small 2}~The subproblem with respect to $a_m(r)$ is expressed as
\begin{equation}
    \min_{a_m(r)} \left[a_m(r) - \tfrac{b_m(r) + c_m(r)}{2} + \tfrac{\breve{\lambda}_2(r) + \theta_m(r) + \phi_m(r)}{2\rho}\right]^2
\end{equation}
and can be easily solved as
\begin{equation} \label{eq:admm_update_z}
    a_m(r) = \tfrac{b_m(r) + c_m(r)}{2} - \tfrac{\breve{\lambda}_2(r) + \theta_m(r) + \phi_m(r)}{2\rho}.
\end{equation}

\subsubsection{Summary of the Algorithm for Problem \eqref{prob:online}} \label{sec:summary_admm}
The proposed algorithm for problem \eqref{prob:online} is summarized as Algorithm 3. Embedding Algorithm 3 into Algorithm 2 to update $\{\mathbf{v}(t),$ $\mathbf{W}(t)\}$, we get an efficient online algorithm for the joint long-term admission control and beamforming problem.%
\begin{table}[!htbp]
  \centering
  \begin{tabular}{rl}
    \toprule
    \multicolumn{2}{l}{{\bf Algorithm 3:} The Low-Complexity Algorithm for Problem \eqref{prob:online} }\\
    \toprule
    1.& Initialize $\mathbf{E},\breve{\mathbf{v}},\mathbf{c}, \boldsymbol{\Omega}, \boldsymbol{\theta}, \boldsymbol{\phi}, \boldsymbol{\epsilon}, \boldsymbol{\delta}, \boldsymbol{\tau}, \boldsymbol{\eta}$;\\
    2.& \texttt{\textbf{repeat}}\quad {\it \textcolor{blue}{(Outer SUM Loop)}} \\
    3.& \quad  $\tilde{\mathbf{v}} \leftarrow \breve{\mathbf{v}}$;\\
    4.& \quad \texttt{\textbf{repeat}}\quad {\it \textcolor{blue}{(Inner ADMM Loop)}} \\
    5.& \qquad  $\breve{\mathbf{W}}(r) \leftarrow \eqref{eq:admm_update_w}$, $\forall~r$;\\
    6.& \qquad  $\{b_m(r), x_m(r), s_m(r)\} \leftarrow \eqref{eq:KKT_xad}$, $\forall~m,r$;\\
    7.& \qquad  Update $\{c_m(r), y_m(r), z_m(r)\}$ similarly, $\forall~m,r$;\\
    8.& \qquad  $\{\mathbf{e}^m(r), \breve{v}_m(r)\} \leftarrow \eqref{eq:admm_update_fs}$, $\forall~m,r$;\\
   9.& \qquad  $a_m(r) \leftarrow \eqref{eq:admm_update_z}$, $\forall~m,r$;\\
   10.& \qquad Update $\boldsymbol{\Omega}, \boldsymbol{\theta}, \boldsymbol{\phi}, \boldsymbol{\epsilon}, \boldsymbol{\delta}, \boldsymbol{\tau}, \boldsymbol{\eta}$ as in \eqref{eq:admm_iteration_lagrangian}; \\
   11.& \quad \texttt{\textbf{until}} the stopping criterion for ADMM is satisfied;\\
   12.& \texttt{\textbf{until}} the stopping criterion for SUM is satisfied;\\
   13.& Output: $\{\mathbf{v}(t), \mathbf{W}(t)\} \leftarrow \{\breve{\mathbf{v}}(1), \breve{\mathbf{W}}(1)\}$.\\
    \bottomrule
  \end{tabular}
\end{table}

Algorithm 3 has two commendable merits. First, each step of the algorithm involves solving a low-dimensional problem. As shown in \eqref{eq:admm_update_w}, \eqref{eq:KKT_xad}, \eqref{eq:admm_update_fs}, and \eqref{eq:admm_update_z}, etc., these problems can be efficiently solved in (semi)-closed form. Consequently, Algorithm 3 has lower complexity than the algorithm solving the high-dimensional problem directly (e.g., the interior-point (IP) algorithm). Second, Algorithm 3 has a separable structure, thus making it applicable to parallel computation. This further improves the efficiency of the algorithm.

For instance, Algorithm 3 can be efficiently implemented in the platform with a multi-core processor. Further, to maximally explore the efficiency of Algorithm 3 in parallel implementation, we assume a sufficient number, e.g., $(J+1)(M+1)$, of cores here. These cores are divided into $(J+1)$ groups, with each group consisting of a main core and $M$ auxiliary (AUX) cores. The $(J+1)$ groups work independently. In particular, group $r$ updates the variables with index $r$, i.e., $\breve{\mathbf{v}}(r), \breve{\mathbf{W}}(r),$ $\mathbf{a}(r), \mathbf{b}(r), \mathbf{c}(r), \cdots$. These tasks are assigned to the $(M+1)$ cores in group $r$ as follows.

$\bullet$ {Main core in group $r$}: updates $\breve{\mathbf{W}}(r)$ and $\boldsymbol{\Omega}(r)$.

$\bullet$ {AUX core $m$ in group $r$}: updates $\{b_m(r), x_m(r), s_m(r)\}$, $\{c_m(r), y_m(r), z_m(r)\}$, $\{\mathbf{e}^{m}(r), \breve{v}_m(r)\}$, $a_m(r)$, and $\{\theta_m(r),$ $\phi_m(r), \alpha_m(r), \beta_m(r), \tau_m(r), \eta_m(r)\}$, $m = 1, 2, \ldots, M$.

\begin{figure}[!htbp]
  \centering
  \includegraphics[width = 8.0cm]{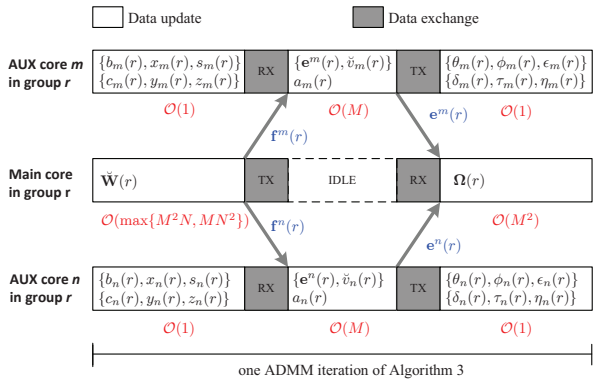}
  \caption{Parallel implementation of Algorithm 3 in platform with a multi-core processor, where ``TX/RX'' denotes ``data transmit/receive''.}
  \label{fig:admm_distribution}
\end{figure}

Fig. \ref{fig:admm_distribution} illustrates the parallel implementation of Algorithm 3. In each iteration, the main core in group $r$ updates $\breve{\mathbf{W}}(r)$ and then computes $\mathbf{F}(r) = [\breve{\mathbf{H}}^\dag(r)\breve{\mathbf{W}}(r),\sigma\mathbf{1}] - \boldsymbol{\Omega}(r)$; at the same time, AUX core $m$ in group $r$ updates $\{a_m(r), x_m(r), s_m(r)\}$ and $\{b_m(r), y_m(r), z_m(r)\}$, $m = 1, 2, \ldots, M$. Next, the main core distributes $\mathbf{f}^m(r)$ to AUX core $m$ through the high-speed on-chip data link. After receiving $\mathbf{f}^m(r)$, AUX core $m$ updates $\{\mathbf{e}^m(r), \breve{v}_m(r)\}$ and $a_m(r)$, and then sends $\mathbf{e}^m(r)$ back to the main core, where the matrix $\mathbf{E}(r)$ is constructed. Using $\mathbf{E}(r)$, the main core updates $\boldsymbol{\Omega}(r)$; at the same time, AUX core $m$ updates $\{\theta_m(r),\phi_m(r), \epsilon_m(r), \delta_m(r),\tau_m(r),$ $\eta_m(r)\}$.

As shown in Fig. \ref{fig:admm_distribution}, the per-iteration complexity of group $r$ is  about $\mathcal{O}(\max\{MN^2, M^2N\})$, which is mainly for updating $\breve{\mathbf{W}}(r)$. Therefore, the per-iteration complexity of Algorithm 3 is $\mathcal{O}((J+1)\cdot\max\{MN^2, M^2N\})$, since we have $(J+1)$ groups. As a contrast, the classic IP method requires the per-iteration complexity of $\mathcal{O}([(J+1)MN]^3)$ to solve problem \eqref{prob:online}. Obviously, compared with the IP method, Algorithm 3 is more efficient and less sensitive to problem dimension, by decomposing problem \eqref{prob:online} into multiple simple low-dimensional subproblems.

\begin{remark} \label{lemma:2}
Following the ADMM framework \eqref{eq:admm_iteration}, problem \eqref{prob:online_uniform_iteration} can be optimally solved in each SUM iteration \cite{Boyd2011J}. Then, referring to the results of Remark \ref{lemma:1}, we claim that every limit point of the iterates generated by Algorithm 3 is a stationary solution of problem \eqref{prob:online}.
\end{remark}



\section{Numerical Simulations} \label{sec:simulation}
Consider a network consisting of one BS and $M = 10$ users, where the BS has $N = 5$ antennas and the users all have single antenna. The BS and users are all within a hexagonal cell, and the distance between adjacent corners is $d_c = 1\,{\rm km}$. The BS is deployed at the center of the cell, while the users are randomly located in the cell. The power budget of BS is $P = 100$, and the noise power is set as $\sigma^2 = 1$.

We employ the block fading channel model --- the channels in each fading block (time slice) are assumed static, while the channels in different fading blocks (time slices) are generated according to Rayleigh distribution. Specifically, the elements of $\mathbf{h}_m(t)$ follow the distribution $\mathcal{CN}(0, \varsigma_m^2(t))$, with $\varsigma_m^2(t) = \varrho_m\cdot[\frac{200}{e_m(t)}]^{3.7}$, where $e_m(t)$ is the distance between BS and user $m$ in time slice $t$, and $\varrho_m$ is the shadowing effect complied to $10\log_{10}\varrho_m \sim \mathcal{N}(0,64)$. We assume the users keep still in the time period, i.e., $e_m(1) = e_m(2) = \cdots = e_m(T)$, $\forall\,m$.

\begin{figure}[tbp]
  \centering
  \includegraphics[width = 7.0cm]{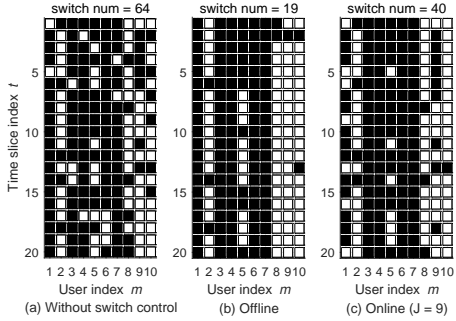}
  \caption{The users' admissible statuses in one simulation trial for $M = 10$, $N = 5$, $\lambda_1 =20$, $\gamma = 1$, and $T = 20$.}
  \label{fig:admissible_status}
\end{figure}

First, we intuitively show how different algorithms balance the flexibility and stability of admissible user set. To this end, we display the users' admissible statuses in one simulation trial in Fig. \ref{fig:admissible_status}, where the horizontal and vertical axes indicate user and time slice, respectively. The black grid means that the user is admissible in the corresponding time slice, while the white grid means inadmissible. In Fig. \ref{fig:admissible_status}(a), we set $\lambda_2 = 0$ and solve problem \eqref{prob:conventional} in different time slices independently; i.e., there is no switching control. As a consequence, the admissible status changes frequently. The total switching number is 64 in this simulation. In Figs. \ref{fig:admissible_status}(b) and \ref{fig:admissible_status}(c), we set $\lambda_2 = 20$ to control the switching frequency of each user's admissible status by the offline and online approaches, respectively. Stabler admissible user sets can be observed in these two subfigures. In addition, the offline approach outperforms the online approach (where we set $J = 9$) in stability control since it uses the actual CSI values in the length-$T$ time period, while the latter only knows the distribution of future CSI because of the practical causality constraint. In this simulation, the switching numbers of the two (offline and online) approaches are 19 and 40, respectively.

Next, we compare the following algorithms in terms of user admission ratio (defined as the ratio of admissible user number to total user number, i.e., $\substack{1 - \frac{\sum_{m,t}\mathcal{I}[v_m(t)]}{MT}}$), user switching frequency (defined as the average switching number per time slice, i.e., $\substack{\frac{\sum_{m,t}|\mathcal{I}[v_m(t+1)] - \mathcal{I}[v_m(t)]|}{T-1}}$), and total system cost (defined as \eqref{obj:original_prob}).

\textcircled{\small 1}~The traditional joint admission control and beamforming algorithm (without switching control), that solves problem \eqref{prob:conventional} in different time slices independently.

\textcircled{\small 2}~The offline approach to problem \eqref{prob:approximate}, i.e., Algorithm 1.

\textcircled{\small 3}~The online approach to problem \eqref{prob:approximate}, i.e., Algorithm 2 with Algorithm 3 embedded in Step 5. Here we set $J = 3$ and $9$ to test the online approach with different sample sizes.

\textcircled{\small 4}~The channel strength based algorithm, that sorts the users in descending order of channel gain, and then admits the users in succession along the sequence. In this algorithm, the number of admissible users is the same as that of the online algorithm with $J = 9$. In each time slice, when the admissible user set is determined, we obtain the transmit beamformers by solving a classic QoS-constrained beamforming problem.

\begin{figure}[tbp]
  \centering
  \includegraphics[width = 7cm]{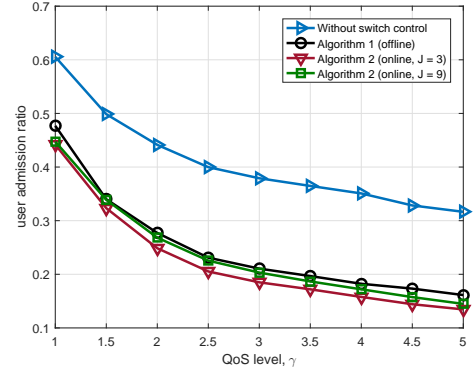}
  \caption{The user admission ratio comparison at different QoS levels ($\gamma$) for $M = 10$, $N = 5$, $\lambda_1 =20$, and $\lambda_2 = 20$.}
  \label{fig:admission_gamma}
\end{figure}

\begin{figure}[tbp]
  \centering
  \includegraphics[width = 7cm]{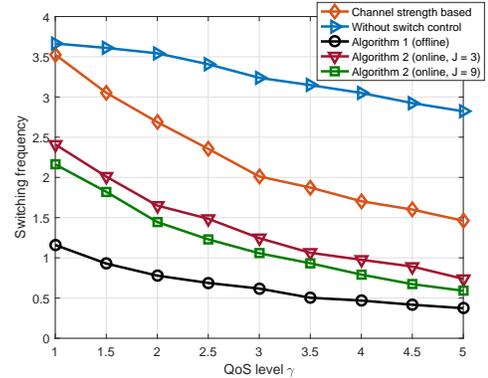}
  \caption{The switching frequency comparison at different QoS levels ($\gamma$) for $M = 10$, $N = 5$, $\lambda_1 =20$, and $\lambda_2 = 20$.}
  \label{fig:switch_gamma}
\end{figure}

\begin{figure}[tbp]
  \centering
  \includegraphics[width = 7cm]{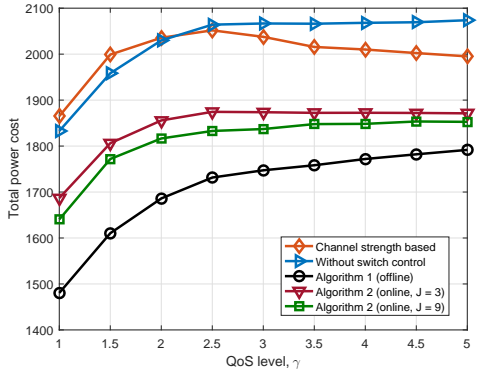}
  \caption{The total power cost comparison at different QoS levels ($\gamma$) for $M = 10$, $N = 5$, $\lambda_1 =20$, and $\lambda_2 = 20$.}
  \label{fig:power_gamma}
\end{figure}

In Figs. \ref{fig:admission_gamma}, \ref{fig:switch_gamma}, and \ref{fig:power_gamma}, we set $\lambda_1 = 20$, $\lambda_2 = 20$, and then compare the algorithms at different QoS levels. As shown in Fig. \ref{fig:admission_gamma}, the admission ratios of the algorithms in comparison all decrease with $\gamma$. The reason is twofold. First, as $\gamma$ increases, the network itself can support less users at their desired QoS levels. Second, since the transmit power for each user increases with QoS level, the network is likely to reject more users as $\gamma$ increases to balance the transmit power. Moreover, compared with the algorithm without switching control, Algorithm 1 and Algorithm 2
take the switching power into consideration. Thus, they further limit the size of user set to avoid unnecessary user switching.

The switching frequency comparison at different QoS levels is displayed in Fig. \ref{fig:switch_gamma}. Generally, the switching frequencies of these algorithms decrease with $\gamma$ due to the shrinking user set. Among them, the traditional algorithm has the highest switching frequency since it selects the users based on instantaneous CSI only. The channel strength based algorithm gets a stabler user set because of our setting of static BS and users. Since the channel strength depends on the distance between BS and user partly, the user closer to BS tends to have stronger channel and the probability of being admissible is higher. Thus, selecting users according to channel strength helps reduce the switching frequency. However, due to the random shadowing factor, the switching performance of this approach is still unsatisfactory. As a contrast, the two proposed (offline and online) algorithms achieve much stabler user set by performing joint optimization on the size of admissible user set, the transmit beamformers, and the switching frequency of each user's admissible status. Further, the offline algorithm outperforms the online approach due to the knowledge of accurate CSI values. For the online algorithm, larger sample size $J$ yields better approximation of the expected power cost in the next time slice, thus improving the stability of user set.

In Fig. \ref{fig:power_gamma}, we show the total power costs of the algorithms. As expected, the offline algorithm performs best, and then the online algorithm follows. Again, increasing $J$ helps reduce the total power cost in the online approach. They outperform the other two algorithms which ignore the switching power in user selection. 
\begin{figure}[tbp]
  \centering
  \includegraphics[width = 7cm]{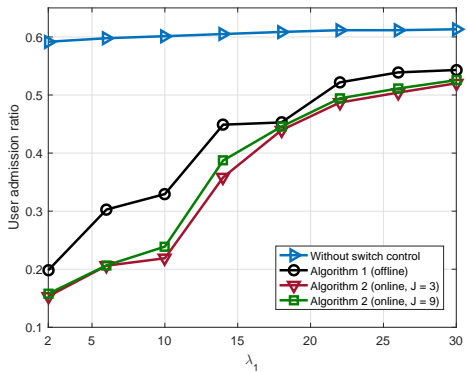}
  \caption{The user admission ratio comparison at different values of $\lambda_1$ for $M = 10$, $N = 5$, $\lambda_2 =20$, and $\gamma = 1$.}
  \label{fig:admission_lambda1}
\end{figure}

\begin{figure}[tbp]
  \centering
  \includegraphics[width = 7cm]{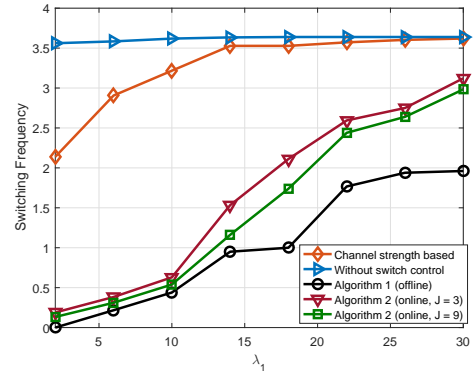}
  \caption{The switching frequency comparison at different values of $\lambda_1$ for $M = 10$, $N = 5$, $\lambda_2 =20$, and $\gamma = 1$.}
  \label{fig:switch_lambda1}
\end{figure}

\begin{figure}[tbp]
  \centering
  \includegraphics[width = 7cm]{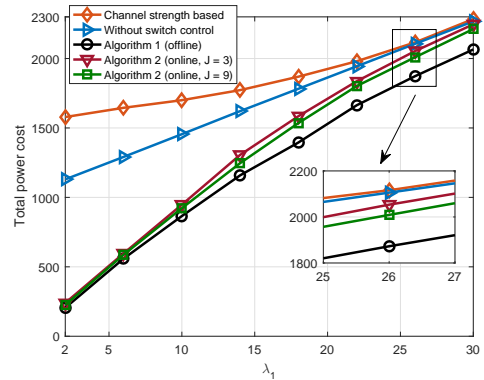}
  \caption{The total power cost comparison at different values of $\lambda_1$ for $M = 10$, $N = 5$, $\lambda_2 =20$, and $\gamma = 1$.}
  \label{fig:power_lambda1}
\end{figure}

In Figs. \ref{fig:admission_lambda1}, \ref{fig:switch_lambda1}, and \ref{fig:power_lambda1}, we fix $\gamma = 1$, $\lambda_2 =20$, and compare the algorithms at different values of $\lambda_1$. In Fig. \ref{fig:admission_lambda1}, we show the user admission ratios of these algorithms. Generally, the admission ratio increases with $\lambda_1$ to alleviate the cost of losing users. In particular, the traditional algorithm, which ignores the issue of user switching and independently solves the admission control and beamforming problems in different time slices, serves the largest number of users. Notice that the admission ratio of this method increases very slowly with $\lambda_1$, implying that under our setting, the network can serve about 60\% of the users at most. For the proposed two algorithms, when $\lambda_1$ is small, the offline algorithm admits more users than the online algorithm. As $\lambda_1$ increases, this difference tends to diminish since for large $\lambda_1$ the network mainly aims to serve as many users as possible. The \emph{stair-like} curves of the offline and online algorithms are mainly due to the fact that we approximate $\mathcal{I}(x)$ by the stair-shape function $1 - \frac{1}{1+\kappa x}$ in problem formulation.

In Fig. \ref{fig:switch_lambda1}, we compare the switching frequencies of these algorithms for different values of $\lambda_1$. As $\lambda_1$ increases, the network emphasizes more on the size of user set, thus loosening the control of switching frequency. Consequently, the switching frequency increases with $\lambda_1$. Among these algorithms, the traditional one suffers from the highest switching frequency. Selecting users according to channel strength can improve the stability of user set to some extent. However, the switching frequency of it is still very high. By optimizing the switching frequency of each user's admissible status, the proposed two algorithms obtain relatively stable user sets. Again, the offline method performs better than the online one in stability control. Increasing $J$ yields stabler BS-user transmission links in the online algorithm.

The total power costs of these algorithms with different $\lambda_1$ are shown in Fig. \ref{fig:power_lambda1}. As expected, the total power cost increases with $\lambda_1$. The channel strength based algorithm performs worst since it selects users based on limited information. The second worst is the traditional algorithm, which neglects the cost of user switching in network management. By applying a system-level optimization strategy, the offline and online algorithms defeat the above two algorithms and achieve lower power cost.

\begin{figure}[tbp]
  \centering
  \includegraphics[width = 7cm]{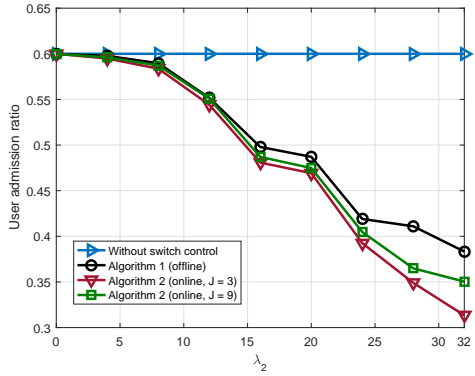}
  \caption{The user admission ratio comparison at different values of $\lambda_2$ for $M = 10$, $N = 5$, $\lambda_1 =20$, and $\gamma = 1$.}
  \label{fig:admission_lambda2}
\end{figure}

\begin{figure}[tbp]
  \centering
  \includegraphics[width = 7cm]{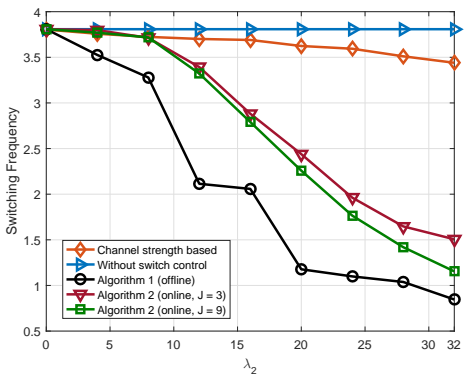}
  \caption{The switching frequency comparison at different values of $\lambda_2$ for $M = 10$, $N = 5$, $\lambda_1 =20$, and $\gamma = 1$.}
  \label{fig:switch_lambda2}
\end{figure}

\begin{figure}[tbp]
  \centering
  \includegraphics[width = 7cm]{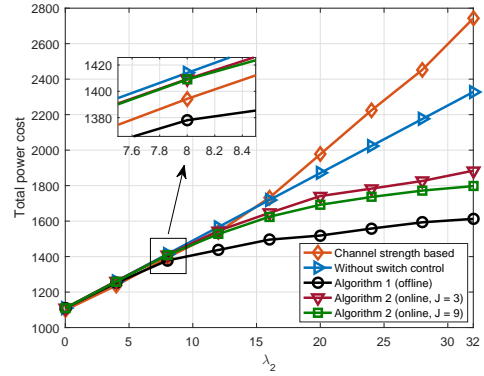}
  \caption{The total power cost comparison at different values of $\lambda_2$ for $M = 10$, $N = 5$, $\lambda_1 =20$, and $\gamma = 1$.}
  \label{fig:power_lambda2}
\end{figure}

In Figs. \ref{fig:admission_lambda2}, \ref{fig:switch_lambda2}, and \ref{fig:power_lambda2}, we fix $\gamma = 1$, $\lambda_1 = 20$, and then compare the algorithms with different values of $\lambda_2$. The user admission ratios and switching frequencies of these algorithms are shown in Figs. \ref{fig:admission_lambda2} and \ref{fig:switch_lambda2}, respectively. Solving problem \eqref{prob:conventional} in different time slices independently, the traditional algorithm is immune to the value of $\lambda_2$. For the other two algorithms, as $\lambda_2$ increases, the switching power accounts for the main part of the total power cost. To effectively control the total power cost, the network must reduce the switching power by limiting the switching frequency of each user (see Fig. \ref{fig:switch_lambda2}). To focus on this, the network loosens the control on the cost of losing users. As a result, the admission ratio decreases with $\lambda_2$ (see Fig. \ref{fig:admission_lambda2}). In this process, the offline algorithm performs better than the online one due to the availability of future CSI.

In Fig. \ref{fig:power_lambda2}, we show the total power cost of these algorithms. In the case of small $\lambda_2$, the influence of switching power can be ignored, and thus these algorithms have similar total power costs. As $\lambda_2$ increases, the switching power
dominates the total poser cost gradually. Then, by performing long-term admission control, the proposed two algorithms show their advantages in controlling the switching power. Finally, they effectively lower the total power cost. Some performance gap can be observed between the offline and online algorithms, since the latter uses the statistics of future CSI in admission control. Increasing the sample size $J$ helps narrow this gap.

\begin{figure}[tp]
  \centering
  \includegraphics[width = 7cm]{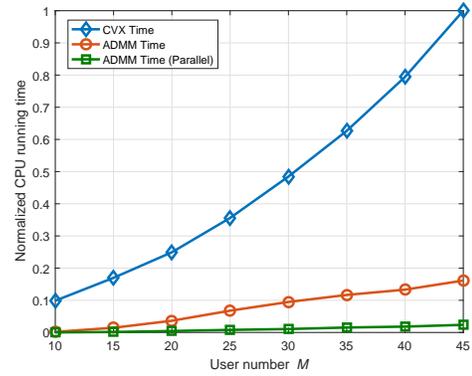}
  \caption{The efficiency comparison at different user numbers $(M)$ for $N = 5$, $\lambda_1 =20$, $\lambda_2 =20$, and $\gamma = 1$.}
  \label{fig:time_user_num}
\end{figure}

In the last simulation, we show the efficiency advantage of the proposed ADMM-based algorithm in solving problem \eqref{prob:online_uniform_iteration}. As a comparison, we solve problem \eqref{prob:online_uniform_iteration} by CVX{\footnote{Usually, the IP method is utilized in CVX to solve the problem \cite{Grant2014M}.}} under the same parameter settings. 
In Fig. \ref{fig:time_user_num}, we show the normalized CPU running times of them. Specifically, in order to simulate the parallel implementation in Fig. \ref{fig:admm_distribution}, we divide the CPU time of ADMM by the number of core groups, and thus obtain the curve labeled ``ADMM time (parallel)''. The results in Fig. \ref{fig:time_user_num} show that the ADMM-based algorithm is more efficient than CVX (or the IP method) in spite of the user number $M$. 

\section{Conclusion} \label{sec:conclusion}
In this paper, we take the stability of user set into account in admission control, and propose a joint long-term admission control and beamforming problem. Our target is to balance the transmit power, the switching power, and the size of admissible user sets by carefully selecting the users and the beamformers in the given time period. We develop two (offline and online) algorithms to handle this challenging NP-hard problem. The offline algorithm requires the knowledge of all CSIs within the time period, and optimizes the variables in different time slices altogether in one shot. To support real-time data transmission, we further design an online framework to solve the admission control and beamforming problem time slice by time slice, utilizing the previous admissible user set, the actual value of current CSI, and the distribution of future CSI. Specifically, in each time slice of the online framework, an ADMM-based algorithm is developed to efficiently solve the joint admission control and beamforming problem. The numerical results have shown that the proposed two algorithms can effectively reduce the network power cost, and the online algorithm is an efficient practical alternative to the offline algorithm in real-time data transmission.

\ifCLASSOPTIONcaptionsoff
  \newpage
\fi

%
%
%
%
%





\begin{thebibliography}{10}

\bibitem{Lin2018C}
J. Lin, Q. Li, and M. Ma, ``Joint long-term admission control and beamforming in downlink MISO networks,'' in {\it Proc. EUSIPCO}, pp. 942--946, Rome, Italy, Sep. 2018.

\bibitem{Ahmed2005J}
M.H. Ahmed, ``Call admission control in wireless networks: A comprehensive survey,'' {\it IEEE Communications Survey \& Tutorials}, vol. 7, no. 1, pp. 49--68, May 2005.

\bibitem{Evangelinakis2010C}
D.I. Evangelinakis, N.D. Sidiropoulos, and A. Swami, ``Joint admission and power control using branch \& bound and gradual admissions,'' in {\it Proc. IEEE SPAWC}, pp. 1--5, Marrakech, June 2010.

\bibitem{Andersin1996J}
M. Andersin, Z. Rosberg, and J. Zander, ``Gradual removals in cellular PCS with constrained power control and noise,'' {\it Wireless Networks}, vol. 2, no. 1, pp. 27--43, Mar. 1996.

\bibitem{Mitliagkas2011J}
I. Mitliagkas, N. D. Sidiropoulos, and A. Swami, ``Joint power and admission control for ad-hoc and cognitive underlay networks: Convex approximation and distributed implementation,'' {\it IEEE Trans. Wireless Communications}, vol. 10, no. 12, pp. 4110--4121, Dec. 2011.

\bibitem{Matskani2008J}
E. Matskani, N.D. Sidiropoulos, Z.-Q. Luo, and L. Tassiulas, ``Convex approximation techniques for joint multiuser downlink beamforming and admission control,'' {\it IEEE Trans. Wireless Communications}, vol. 7, no. 7, pp. 2682--2693, Jul. 2008.

\bibitem{Liu2013J}
Y.-F. Liu, Y.-H. Dai and Z.-Q. Luo, ``Joint power and admission control via linear programming deflation,'' {\it IEEE Trans. Signal Processing}, vol. 61, no. 6, pp. 1327--1388, June 2013.

\bibitem{Liu2015J}
Y.-F. Liu, Y.-H. Dai and S. Ma, ``Joint power and admission control: Non-convex $\ell_q$ approximation and an effective polynomial time deflation approach,'' {\it IEEE Trans. Signal Processing}, vol. 63, no. 14, pp. 3641--3656, Jul. 2015.

\bibitem{Lin2017J}
J. Lin, R. Zhao, Q. Li, H. Shao, and W.-Q. Wang, ``Joint base station activation, user admission control and beamforming in downlink green networks,'' \textit{Digital Signal Processing}, vol. 68, no. 9, pp.~182--191, Sep. 2017.

\bibitem{Wai2012C}
H.-T. Wai and W.-K. Ma, ``A decentralized method for joint admission control and beamforming in coordinated multicell downlink,'' in {\it Proc. IEEE ASILOMAR}, pp.~559--563, Pacific Grove, USA, Nov. 2012.

\bibitem{Lai2016J}
W.-S. Lai, T.-H. Chang, and T.-S. Lee, ``Joint power and admission control for spectral and energy efficiency maximization in heterogeneous OFDMA networks'' {\it IEEE Trans. Wireless Communications}, vol. 15, no. 5, pp. 3531--3547, May 2016.

\bibitem{Kuang2012C}
Q. Kuang, J. Speidel, and H. Droste, ``Joint base-station association, channel assignment, beamforming and power control in heterogeneous networks,'' in {\it Proc. IEEE VTC-Spring}, pp. 1--5, Yokuhama, May 2012.

\bibitem{Azam2016J}
M. Azam, M. Ahmad, M. Naeem, M. Iqbal, A.S. Khwaja, A. Anpalagan, and S. Qaisar, ``Joint admission control, mode selection, and power allocation in D2D communication systems,'' \textit{IEEE Trans. Vehicular Technology}, vol. 65, no. 9, pp. 7322--7333, Sep. 2016.

\bibitem{Liu2012C}
P. Liu, C. Hu, T. Peng, and W. Wang, ``Distributed cooperative admission and power control for device-to-device links with QoS protection in cognitive heterogeneous network,'' in {\it Proc. CHINACOM}, pp.~712--716, Kunming, China, Aug. 2012.

\bibitem{Monemi2015J}
M. Monemi, M. Rasti, and E. Hossain, ``On joint power and admission control in underlay cellular cognitive radio networks,'' \textit{IEEE Trans. Wireless Communications}, vol. 14, no. 1, pp. 265--278, Jan. 2015.

\bibitem{Zander1992J1}
J. Zander, ``Performance of optimum transmitter power control in cellular radio systems,'' \textit{IEEE Trans. Vehicular Technology}, vol. 41, no. 1, pp.~57--62, Jan. 1992.

\bibitem{Lin2017C1}
J. Lin and R. Zhao, ``An distributed deflation algorithm for joint admission control and beamforming in multi-user max-min fairness networks,'' in {\it Proc. the 23rd Asia-Pacific Conference on Communications (APCC)}, pp.~1--6, Perth, Australia, Dec. 2017.

\bibitem{Lin2019J}
J. Lin, C. Gu, J. Yang, Q. Li, and W.-Q Wang, ``Joint admission control and beamforming in max-min fairness networks,'' {\it IET Communications}, vol. 13, no. 13, pp. 1953--1961, Aug. 2019.

\bibitem{Chen2011J}
T. Chen, Y. Yang, H. Zhang, and H. Kim, ``Network energy saving technologies for green wireless access networks,'' {\it IEEE Wireless Communications}, vol. 18, no. 5, pp. 30--38, May 2011.

\bibitem{Han2011J}
C. Han, T. Harrold, S. Armour, et al., ``Green radio: Radio techniques to enable energy-efficient wireless networks,'' {\it IEEE Communications Magazine}, vol. 49, no. 6, pp. 46--54, May 2011.

\bibitem{Hasan2011J}
Z. Hasan, H. Boostanimehr, and V.K. Bhargava, ``Green cellular networks: A survey, some research issues and challenges,'' {\it IEEE Communications Surveys \& Tutorials}, vol. 13, no. 4, pp. 524--540, Fourth Quarter 2011.

\bibitem{Simsek2015C}
M. Simesk, M. Bennis, and I. G\"{u}ven\c{c}, ``Context-aware mobility management in HetNets: A reinforcement learning approach,'' in {\it Proc, IEEE WCNC}, pp. 1536--1541, New Orleans, LA, USA, Mar. 2015.

\bibitem{Pan2012C}
J. Pan and W. Zhang, ``An MDP-based handover decision algorithm in hierarchical LTE networks,'' in {\it Proc. IEEE VTC-Fall}, pp. 1--5, Quebec City, Sep. 2012.

\bibitem{Lin2017C2}
J. Lin, Q. Li, and H. Deng, ``An online algorithm for joint long-term BS activation and beamforming in green downlink MISO networks,'' in {\it Proc. EUSIPCO}, pp. 435--439, Kos island, Greece, Aug. 2017.

\bibitem{Sun2013C}
R. Sun, H. Baligh and Z.-Q. Luo, ``Long-term transmit point association for coordinated multipoint transmission by stochastic optimization,'' in {\it Proc. IEEE SPAWC}, pp. 330--334, Darmstadt, June 2013.

\bibitem{Yu2016J}
N. Yu, Y. Miao, L. Mu, H. Du, H. Huang, and X. Jia, ``Minimizing energy cost by dynamic switching on/off base stations in cellular networks,'' {\it IEEE Trans. Wireless Communications}, vol. 15, no. 11, pp. 7457--7469, Nov. 2016.

\bibitem{Chen2017J}
Q. Chen, D. Kang, Y. He, T.-H. Chang, and Y.-F. Liu, ``Joint power and admission control based on channel distribution information: A novel two-timescale approach,'' {\it IEEE Signal Processing Letters}, vol. 24, no. 2, pp. 196--200, Feb. 2017.


\bibitem{Meisam2013J}
M. Razaviyayn, M. Hong, and Z.-Q. Luo, ``A unified convergence analysis of block successive minimization methods for nonsmooth optimization,'' {\it SIAM Journal on Optimization}, vol. 23, no. 2, pp. 1126--1153, June 2013.

\bibitem{Shapiro2006J}
A. Shapiro, ``On complexity of mutistage stochastic programs,'' {\it Operations Research Letters}, vol. 34, no. 1, pp. 1--8, Jan. 2006.

\bibitem{Boyd2011J}
S. Boyd, N. Parikh, E. Chu, B. Peleato, and J. Eckstein, ``Distributed optimization and statistical learning via the alternating direction method of multipliers,'' {\it Foundations and Trends in Machine Learning}, vol. 3, no. 1, pp. 1--122, 2011.

\bibitem{Shen2012J}
C. Shen, T.-H. Chang, K.-Y. Wang, Z. Qiu, and C.-Y. Chi, ``Distributed robust multicell coordinated beamforming with imperfect CSI: an ADMM approach,'' {\it IEEE Trans. Signal Processing}, vol. 60, no. 6, pp. 2988--3003, June 2012.


\bibitem{Biglieri1998J}
E. Biglieri, J. Proakis, and S. Shamai, ``Fading channels: Information theoretic and communication aspects,'' {\it IEEE Trans. Information Theory}, vol. 44, no. 6, pp. 2619--2692, Oct. 1998.



\bibitem{Grant2014M}
M. Grant, S. Boyd, ``The CVX users' guide,'' {\it CVX Research, Inc.}, Oct. 24, 2014 [On-line]. http://cvxr.com/cvx/.

\bibitem{Hestenses1969J}
M. Hestenes, ``Multiplier and gradient methods,'' {\it Journal of Optimization Theory and Applications}, vol. 4, pp. 303--320, 1969.

\bibitem{Boyd2004M}
S. Boyd, and L. Vandenberghe, {\it Convex Optimization}, Cambridge, UK: Cambridge University Press, 2004.

\bibitem{Lin2014C}
J. Lin, Y. Li, and Q. Peng, ``Joint power allocation, base station assignment and beamformer design for an uplink SIMO heterogeneous network,'' in {\it Proc. IEEE ICASSP}, pp. 434--438, Florence, May 2014.



\end{thebibliography}
\end{document}